\documentclass [prr,reprint,notitlepage,superscriptaddress,twocolumn,floatfix,nofootinbib] {revtex4-1}
\usepackage{amsmath}
\usepackage{graphicx}
\usepackage{lmodern}
\usepackage{amsmath}

\usepackage{color}
\usepackage{bm}
\usepackage{hyperref}
\usepackage{amssymb}

\newcommand{\beq} {\begin{equation}}
\newcommand{\eeq} {\end{equation}}
\newcommand{\bea} {\begin{eqnarray}}
\newcommand{\eea} {\end{eqnarray}}
\newcommand{\be} {\begin{equation}}
\newcommand{\ee} {\end{equation}}
\renewcommand{\(}{\left(}
\renewcommand{\)}{\right)}
\renewcommand{\[}{\left[}
\renewcommand{\]}{\right]}

\DeclareMathOperator{\sgn}{sgn}

\DeclareMathOperator{\Imm}{Im}
\AtBeginDocument{\renewcommand{\natexlab}[1]{#1}}% <--- the fix

\begin{document}

\title {Quantum Phase Transition  in the Yukawa-SYK Model}
\author{Yuxuan Wang}
\email{yuxuan.wang@ufl.edu}
\affiliation{Department of Physics, University of Florida, Gainesville, FL 32601, USA}
\author{Andrey V. Chubukov}
\email{achubuko@umn.edu}
\affiliation{School of Physics and Astronomy and William I. Fine Theoretical Physics Institute, University of Minnesota, Minneapolis, MN 55455, USA}
\date{\today}
\begin{abstract}
We study the quantum phase transition upon varying the fermionic density  $\nu$ in a solvable model with random Yukawa interactions between $N$ bosons and $M$ fermions, dubbed the Yukawa-SYK model.   We show that there are two distinct phases in the model: an incompressible state with gapped excitations and  an exotic quantum-critical,  non-Fermi liquid state with exponents varying with $\nu$.   We show analytically and numerically that the quantum phase transition between these two states is first-order, as for some range of $\nu$ the NFL state has a negative compressibility. In the limit $N/M\to \infty$ the first-order transition gets weaker and asymptotically becomes second-order, with an exotic quantum-critical behavior. We show that  fermions and bosons display highly unconventional spectral behavior in the transition region.
\end{abstract}
\maketitle

\section{Introduction}
The non-Fermi liquid (NFL)
 is  one of the most fascinating phenomena in modern condensed matter physics. It violates
   the fundamental
   Landau    paradigm
 that quasiparticles  must  become weakly damped at low enough energies.
  The key feature of a NFL
   is  a power-law    form of the fermionic self-energy,
   $\Sigma (\omega) \propto \omega^x$ with $x <1$,
    which leads to
the vanishing of the quasiparticle  residue at $\omega =0$.
 The NFL
 behavior has
 been observed
      in
       quite a few unconventional superconducting materials~\cite{hussey-mackenzie-1998,stewart-2001,shibauchi-carrington-matsuda-2014,
       keimer-kivelson-norman-2015,varma-littlewood-1989}.
It is widely believed to develop for itinerant  fermions near a density-wave or $q=0$ instabilities in either spin or charge channels~\cite{hertz-1976,*millis-1993,oganesyan-kivelson-fradkin-2001,*metzner-2003,*metzner_2006,*rech-pepin-2006,
          *maslov,*lawler-fradkin-2007,*khvesh2006,acf,*acs,vekhter2004,*wang-abanov-2016,wolfle-rmp-2007,senthil-2008,
          metlitski-sachdev-2010-1,*metlitski-sachdev-2010-2,jiang-fisher-2013,
          torroba-2015,sslee-2009,*lee-review-2018,lederer-2015,*schattner-2016,*xu-sun-2017,*klein-2020,xu-klein-2020}, and in systems with
         fermionic excitations coupled to emergent gauge fields~\cite{lee-nagaosa-1992,hlr-1993,aim,*Nayak1994,*polchinski-1994,kim-lee-wen-1995,*kim-lee-wen-stamp-1995,
         jain-anderson-2009}, such as in quantum spin liquids and half-filled Landau levels.

The theoretical understanding of  a NFL remains a challenge.
 Most of earlier studies of NFLs considered itinerant fermions coupled to soft bosonic modes near a quantum-critical point (QCP). These models show non-trivial NFL behavior at the one-loop order, however in most cases the loop expansion    is
   not controlled because of logarithmic singularities,
    even in the large $N$ limit,
   ~\cite{sslee-2009,*lee-review-2018,metlitski-sachdev-2010-1,*metlitski-sachdev-2010-2}, and
     one has to introduce additional
      modifications to the model~\cite{mross-mcgreevy-2010,*metlitski-mross-2015,dalidovich-lee-2013,raghu-torroba-2019}, e.g., dimension regularization
      or matrix large-$N$, to keep the calculations under control.
 Another route to NFL, which has emerged recently~\cite{gu-2017,balents-2017, debanjan-prx,*chowdhury-berg-2019,xu-2019, franz-2019}, explores Sachdev-Ye-Kitaev(SYK)-type
   models~\cite{SY, K, sachdev-2015, maldacena-2016, SYK3, atland2019, gu-2020}. These models describe randomly  interacting  fermions in a quantum dot. The advantage of SYK model is that it is exactly solvable in the large-$N$ limit and displays NFL behavior with a particular fractional exponent  $x=1/2$ for the self-energy.
 Besides, the SYK model has a hidden holographic connection to quantum black holes~\cite{K,sachdev-2015,sachdev-2015,davison-2017} and in this respect is a simple prototypical model for both NFL and quantum gravity.

In this communication, we consider the
generalization of the SYK model, the Yukawa-SYK (Y-SYK) model~\cite{wang-2020,esterlis-2019,esterlis-2019-2}, in which
 $M$ flavors of dispersion-less fermions in a quantum dot randomly interact with $N$
    flavors of  massive bosons, e.g., optical phonons or gapped collective spin or  charge fluctuations.   The interest to this model has been triggered by
  its rich and unconventional physics,  and by recent experimental discoveries
     of strongly correlated behavior
      in flat band systems like
      magic angle twisted bilayer graphene~\cite{cao-2018,cao-2018-2} and $d_{xy}$ band in  Fe-based superconductors~\cite{coldea-2018}.
      The Y-SYK model has been earlier studied
      at half-filling~\cite{wang-2020,esterlis-2019,pan-2020}.
        It was shown that
      the
      interaction
       ``self-tunes" the system  into a NFL, quantum-critical (QC) regime, despite that a bare bosonic mass is finite.This QC regime may in turn become unstable towards non-BCS superconductivity.~\cite{wang-2020,esterlis-2019,esterlis-2019-2,cenke-syk-1,cenke-syk-2,patel-kim-2018, zhao-prx-2019, chowdhury-berg-2019,gnezdilov-2019,wang-kamenev-2020} Like the SYK model, the Y-SYK model also saturates the upper bound for the onset rate of quantum chaos~\cite{kim-cao-altman-2019}, indicating {the existence a classical holographic dual.}

We report the results of on the Y-SYK model  away from half-filling,
 at   fermionic density
  $\nu \neq1/2$.  For small deviations from $\nu=1/2$, we analytically obtain low-energy forms of the fermionic and bosonic Green's functions with NFL exponents and show that
 the fermionic self-energy
  and the spectral function
 become asymmetric in frequency.
  At
 $\nu =1$,   we show that  fermions form an incompressible state and bosons remain gapped.
We then focus on the quantum phase transition between the compressible NFL state
and the incompressible state.  We show both  analytically and numerically that the phase transition is generally first-order because the chemical potential is a non-monotonic function of $\nu$, and
the compressibility $d\nu/d\mu <0$ for a range of $\nu$. We argue that this is
due to robust low-energy properties of the Y-SYK model. In the transition region the fermionic  and bosonic
    spectral functions displays a peculiar precursor behavior~\cite{kotliar-2006}.
In the particular limit, where the number of
 bosonic flavors well exceeds
 the number of fermionic ones,
  non-monotonicity disappears and the transition becomes second-order.
 Even in this case, bosons displays a highly non-trivial ``gap filling" behavior:
the bosonic mass gap
  remains finite on both sides of the transition, but on the NFL side of the transition the spectral weight
      develops       around zero energy,
       and the width of the range, where this happens, increases as the system moves deeper into the NFL region.
    Taken together, these results reveal rich and universal behavior of zero-dimensional quantum-critical NFL systems.

  Some features of the Y-SYK model, like the asymmetry of fermionic self-energy $\Sigma (i\omega)$,
   are similar to those of the complex SYK model~\cite{gps-2001, gu-2020}. Recent numerical results~\cite{azeyanagi-2018,patel-sachdev-2019} for this model suggest that it
      may also undergo a first-order quantum phase transition between a NFL state and an insulating state.  However,
the analytical understanding of that transition is still lacking.
   In particular it remains unclear whether the first-order transition is a universal property of the complex SYK model, or it depends on  non-universal aspects of
   the system behavior
    at larger frequencies.

\section{The model} The  Y-SYK model  describes $M$ flavors of dispersion-less fermions,  randomly coupled to $N$ flavors of
bosons, each  with a finite mass $m_0$.   The dynamics of the model on the Matsubara axis is described by
the Lagrangian
\begin{align}
L =&  \sum_{i,j=1}^{M}\[ c^\dagger_{i}\(\partial_\tau-\mu\)c_{i}\]+\sum_{\alpha=1}^N\[\frac{1}{2}(\partial_\tau\phi_{\alpha})^2
+\frac{m_0^2}{2}\phi_{\alpha}^2\]\nonumber\\
&+\frac{i}{\sqrt{MN}}\sum_{ij\alpha} t_{ij}^{\alpha} c^\dagger_i c_j \phi_{\alpha},~~~~\(t_{ij}^{\alpha}=-t_{ji}^{\alpha}\).
\label{eq:1}
\end{align}
where $\{i,j\}$  are fermion flavor indices, $\{\alpha\}$
  labels the bosons,
  and $\mu$ is the chemical potential. {We have kept the spin indices implicit.} In an open system $\mu$ is a free (input) parameter, while  in a closed system its value is  set by the fermionic density per flavor $\nu\equiv \langle c^{\dagger}_i c_i\rangle$.
The Yukawa fermion-boson coupling is assumed to be random: $\langle t_{ij}^\alpha\rangle=0$, $\langle t_{ij}^\alpha t_{kl}^\beta\rangle=\(\delta_{ik}\delta_{jl}+\delta_{il}\delta_{jk}\)\delta^{\alpha\beta} \omega_0^3$.
 We assume $\omega_0$ to be positive.
 We have chosen the Yukawa coupling to be imaginary, such that the effective interaction in the Cooper channel is repulsive.~\cite{esterlis-2019,wang-2020}
   The model has an exact particle-hole symmetry, under which $\mu\to-\mu$.     For definiteness we set
       $\mu>0$. Previous studies have focused on the system at the half filling $\nu=1/2$, in which
        case
        $\mu=0$.

 The model has three energy scales:  the bare mass of a boson $m_0$, the strength of the Yukawa coupling $\omega_0$, and the chemical potential $\mu$. We will focus on the ``weak-coupling limit" $\omega_0\ll m_0$.  We will see that in this limit there are only two relevant energies, $\mu$ and
 $\omega_F = \omega_0^3/m_0^2$.  We emphasize that
  already at weak coupling, the system behavior at low energies is highly non-perturbative and includes self-tuned criticality and NFL.
 We work at $T=0$ and take both $M$ and $N$ as large numbers, but keep the ratio $N/M$ is a
  parameter.

 At the bare level bosons are gapped, and fermions are free dispersionless quasiparticles. Our goal is to find the fully dressed bosonic and fermionic propagators $G^{-1}(i\omega) =
i\omega+\Sigma(i\omega)+\mu$ and $D^{-1}(i\Omega) =  \Omega^2+ \Pi(i\Omega)+m_0^2$.  We extended results of earlier analysis at half-filling~\cite{wang-2020,esterlis-2019,esterlis-2019-2} to $\mu  \neq 0$
      and found that for $M,N \gg 1$ the fermionic and bosonic self-energies are expressed self-consistently via the Schwinger-Dyson equations
\begin{align}
     \Pi (i\Omega) =& \frac{2M}{N}\omega_0^3\int_\omega
     G(i\omega-i\Omega/2) G(i\omega+i\Omega/2)\nonumber\\
    \Sigma(i\omega) =& -\omega_0^3
    \int_\omega
     D(i\Omega) G(i\omega-i\Omega).
     \label{eq:dyson}
\end{align}
where $\int_\omega \equiv \int d\omega/(2\pi)$.

 We first show that the system behavior is qualitatively different at larger $\mu$ and at smaller $\mu$, and  then consider the phase transition  between the two phases by tuning $\mu$($\nu$) in an open (closed) system.

 \section{Incompressible gapped phase at large $\mu$}
 The point of departure for the analysis at large $\mu$ is the observation that within a direct perturbative expansion
 the bosonic polarization
  \begin{align}
   \Pi (i\Omega) \sim
    \int_\omega
    \frac{1}{i(\omega +\Omega/2)+\mu}\frac{1}{i(\omega+\Omega/2)+\mu} =0
   \label{eq:11}
 \end{align}
 because the poles of the integrand are  in the same frequency half-plane.
Using bare $D(i\Omega) = 1/(\Omega^2 + m^2_0)$ we obtain for
 the fermionic self-energy
 \begin{align}
\Sigma (i\omega)  = -   \int_\Omega
\frac{\omega_0^3}{\Omega^2+ {m_0^2}}\frac{1}{i(\omega-\Omega)+
\mu} \approx - \omega_F/2.
 \label{eq:2}
 \end{align}
 Substituting this into
 (\ref{eq:dyson}), we find
 $G (i\omega) =1/(i \omega  + (\mu  -
 \mu^*))$, where $\mu^* =\omega_F/2$.
The self-energy comes from low-energy fermions and remains the same
 if we compute it self-consistently.
  Similarly,
   $\Pi (i\Omega)$   still vanishes if we re-evaluate it with dressed fermionic Green's functions~\cite{1footnote}.
 This self-consistent approach is valid as long as fermions are gapped, i.e.,  $\mu >\mu^*
 $.  At smaller $\mu$, such solution does not exist, as one can easily verify.
  Because fermionic energies are all negative,  the filling $\nu=1$ independent on $\mu > \mu^*$,
    hence this phase is incompressible (the compressibility $d\nu/d\mu =0$).

 \section{NFL phase at smaller $\mu$}~~~
At  half-filling ($\nu =1/2$, $\mu =0$), previous studies have found that
$\Pi (i \Omega) + m_0^2 \propto  |\Omega|^{1-2x_0}$ and
 $\Sigma (i \omega) +\mu  \propto i |\omega|^{x_0} \sgn \omega$,
  where $x_0$ is a function of $N/M$, ranging from $x_0 = 1/2 - (M/2\pi N)^{1/2}$
  at $N/M \to \infty$ to $x_0=0$ at $N/M \to 0$ ($x_0 = 0.16$ for $N=M$, see Eq.\ \eqref{eq:24a} below).
   This  NFL behavior holds at small frequencies  for {\it any} $m_0$ and {\it any} non-zero $\omega_F$.
   Note that      $\Pi (0) = -m^2_0$,
      i.e.,
       the dressed bosonic mass vanishes. This implies that the system self-tunes to quantum critical regime, despite that the bare mass is large compared to the strength of the interaction ($m_0 \gg \omega_0$).

For a nonzero
$\mu$,  we find that
bosons remain massless and fermions  retain NFL behavoir, but fermionic self-energy becomes an asymmetric function of $\omega$.
 Specifically, for
  $\Omega, \omega \ll \omega_F$,
\begin{align}
& \Sigma(i\omega)   + \mu \equiv \tilde\Sigma(i\omega)    = \omega_f \left|\frac{\omega}{\omega_f}\right|^x \left(i \sgn(\omega) + \alpha\right)
 \nonumber\\
& \Pi(i\Omega) + m_0^2 \equiv \tilde\Pi(i\Omega)
= \beta m_0^2 \left|\frac\Omega{\omega_f}\right|^{1-2x},
\label{eq:asz}
\end{align}
where $\alpha$ parametrizes
spectral asymmetry~\cite{gps-2001}
 and $\omega_f$ is the NFL energy scale,
   below which $|\Sigma(i\omega)|
    < \omega$.
Altogether we have four dimensionless parameters: $x$, $\alpha$, $\beta$, and $\omega_f/\omega_F$.
 Substituting these forms into  Eq.~\eqref{eq:dyson} and matching the power-law parts $\tilde\Sigma$ and $\tilde \Pi$,
we obtain two equations:
 (see Appendix \ref{sec:nfl})
\begin{align}
\label{eq:24a}
\left(1-\alpha^2\right)\frac{1+\sec \pi x}{1/x-2} +\frac{2\alpha^2}{1/x-2} =& (1+\alpha^2)\frac{N}{2M},\\
\label{eq:24}
\frac{\omega_F}{4\pi\beta\omega_f(1+\alpha^2)}\frac{\Gamma^2(-x)}{\Gamma(-2x)}=&-1.
\end{align}
These relations are exact
 as relevant fermionic and bosonic frequencies in \eqref{eq:dyson} are comparable to
external $\omega, \Omega$, which we
  set to be much smaller than $\omega_f$.
Note that the matching the real and  the imaginary parts of ${\tilde \Sigma}(i\omega)$ gives the same equation. Physical values of the exponent $x$ in (\ref{eq:24a}, \ref{eq:24})  are $x \leq 1/2$. A larger $x$ would lead to negative $\beta\omega_f$, which violates the unitarity of the theory. For $x \to 1/2$, $\alpha \to 1$ ($\alpha-1 \approx \pi (1/2-x)$). At half-filling, $\alpha =0$ and $x=x_0$ is a function of $M/N$ ($x_0 \approx 0.16$ at $M/N=1$).
The two other conditions are
$\Sigma (0) =- \mu$ and $\Pi (0) = -m^2_0$.
 Using Eq.~(\ref{eq:dyson}) we obtain
\begin{align}
\frac{\mu}{\omega_F} =\int_\omega\frac{1}{i\omega+\tilde\Sigma(\omega)}\frac{1}{\omega^2+\tilde\Pi(\omega)} ,
\frac{1}{\omega_F} =\int_\omega\frac{-{2M}/{N}}{\(i\omega+\tilde\Sigma(\omega)\)^2}.
\label{ch_1}
\end{align}
Substituting $\Sigma (i\omega)$ and $\Pi (i\Omega)$ from (\ref{eq:asz}) we obtain
\be
\beta = F_1 (x),~~ \alpha = \frac{\mu}{\omega_F}  F_2(x)
\label{eq:mu}.
\ee
The functions $F_{1,2} (x)$ are regular $\mathcal{O}(1)$ functions of the argument, and
$F_1(1/2) = 1$, $F_2 (1/2) =1/2$.
 We
 present
 them in Appendix \ref{sec:nSC}. The relations (\ref{eq:mu}) are not exact, because relevant $\omega$ in (\ref{ch_1})
are of order $\omega_f$. For these $\omega$, Eqs. (\ref{eq:asz}) are valid up to corrections $\mathcal{O}(1)$, as the  forms of $\Sigma (i\omega)$ and $\Pi (i\Omega)$ change at $\omega, \Omega > \omega_f$:
 the bosonic self-energy gradually decreases
and the fermionic $\Sigma (\omega)$ acquires a Fermi liquid form.~\cite{2footnote} Nevertheless, we find that the
relations $F_1(1/2) = 1$
 and $F_2(1/2)
 = 1/2$ are actually exact (see Appendix \ref{sec:nSC}), i.e., at $x=1/2$,  $\mu = \mu^*=\omega_F/2$.

While $\alpha$ cannot be universally expressed via the chemical potential, it can be exactly expressed via the density $\nu$. Using the Luttinger relation between $\nu$ and  properly regularized $\int G(i\omega) d\omega$, we obtain (see Appendix \ref{sec:lutt})
 \be
\nu =\frac{1}{2} + \frac{\tan^{-1} \alpha}{\pi} + \frac{x}{2\sin(\pi x)}\frac{2\alpha}{1+\alpha^2}.
\label{eq:lutt}
 \ee
The same relation holds for the complex SYK model~\cite{gps-2001, gu-2020}.
 From Eqs. (\ref{eq:lutt}) and (\ref{eq:24}) we see that as $x\to 1/2$
   and $\alpha \to 1$, the filling $\nu$ approaches $1$ and $\omega_f$ tends to zero.
   This implies that the range of  NFL
     behavior
     vanishes
     at $\nu \to 1$.

\begin{figure}
\includegraphics[width=\columnwidth]{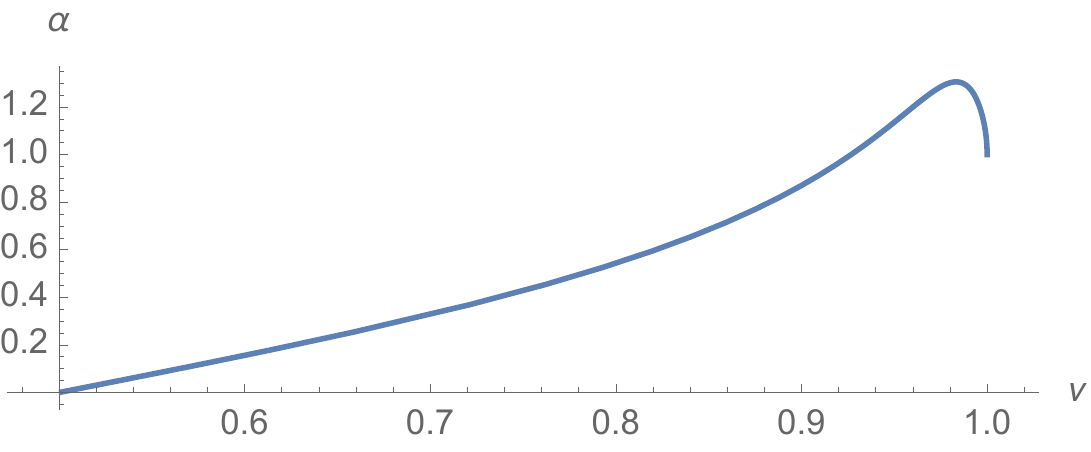}
\includegraphics[width=\columnwidth]{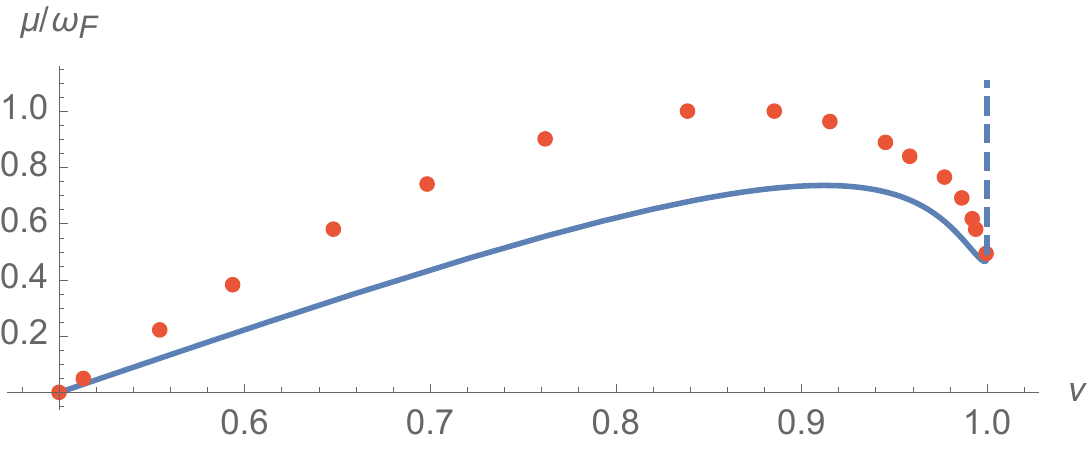}
\caption{
Upper panel: the
 dependence of the spectral asymmetry parameter $\alpha$
   on the
   filling $\nu$.
     Lower panel: the qualitative (solid line) and numerical (red dots) dependence of the  chemical potential $\mu$ on the filling $\nu$. The solid line was
        obtained by using the low-energy form \eqref{eq:asz} for the self-energies for all frequencies.  The vertical dashed line in the lower panel represents the incompressible phase.  We set $N =M$, in which case $x_0 = 0.16$. The solid line actually has a minute dip at $\nu \approx 0.98$ (see Appendices \ref{sec:nSC} and \ref{sec:lutt}).
It is not present in the full numerical solution and  likely is an artifact of using Eq. (5) at all energies.}
\label{fig:1}
\end{figure}

Combining Eqs.~(\ref{eq:24a}, \ref{eq:mu} ,\ref{eq:lutt}),
 we obtain $\alpha$ and $\mu$ as functions of the filling $\nu$. We plot these two functions
    for $M=N$    in Fig.~\ref{fig:1}.
  We see that both $\alpha$ and $\mu$ are non-monotonic functions of $\nu$, and there is a range of $\nu$ where the compressibility $d\nu/d\mu$ is negative.
  The relation $\alpha (\nu)$ is exact, the
   other one, $\mu (\nu)$,    is approximate, as  to get it we used
      Eq.~(\ref{eq:mu}).
 To verify that the nonmonotonic behavior of $\mu$ is not the artifact of our approximation, we iteratively solved the  nonlinear integral equations \eqref{eq:dyson} for $\Sigma(i\omega)$ and $\Pi(i\Omega)$  for all frequencies, using the analytical power-law forms in \eqref{eq:asz} as an input.
    We  show the numerical results~\cite{code} for $\mu(\nu)$ for $M=N$
     in Fig.~\ref{fig:1}.
      We see that the non-monotonic behavior persists.

\section{Quantum phase transition}~~~  The existence of a range of densities, where
${\partial{\nu}}/{\partial{\mu}}$ is negative,
 implies that the NFL solution is unstable, and the transition between the NFL and the insulating state
 must be first order.
In an open system, there is a discontinuous transition  to the incompressible phase~\cite{wiki} at some critical $\mu_c$.
In a closed system, there is a phase coexistence region, in which the system displays simultaneously gap features  from the incompressible state
 and NFL features at small frequencies (see  Fig. \ref{fig:spec}(a)).
 This resembles the ``gap filling" behavior near a Mott transition at $T=0$.~\cite{kotliar-2006}

\begin{figure}[t]
\includegraphics[width=\columnwidth]{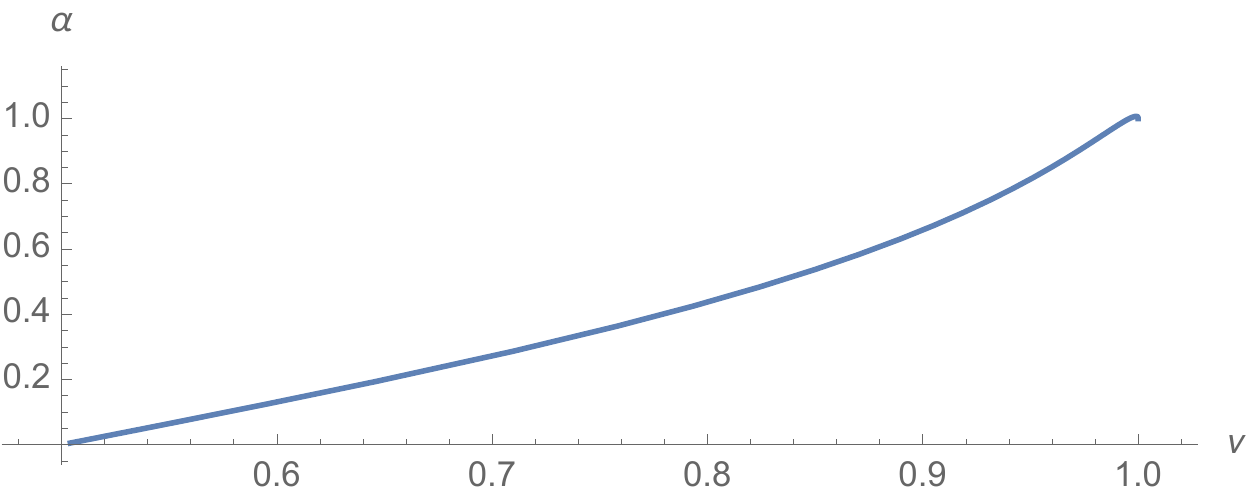}
\includegraphics[width=\columnwidth]{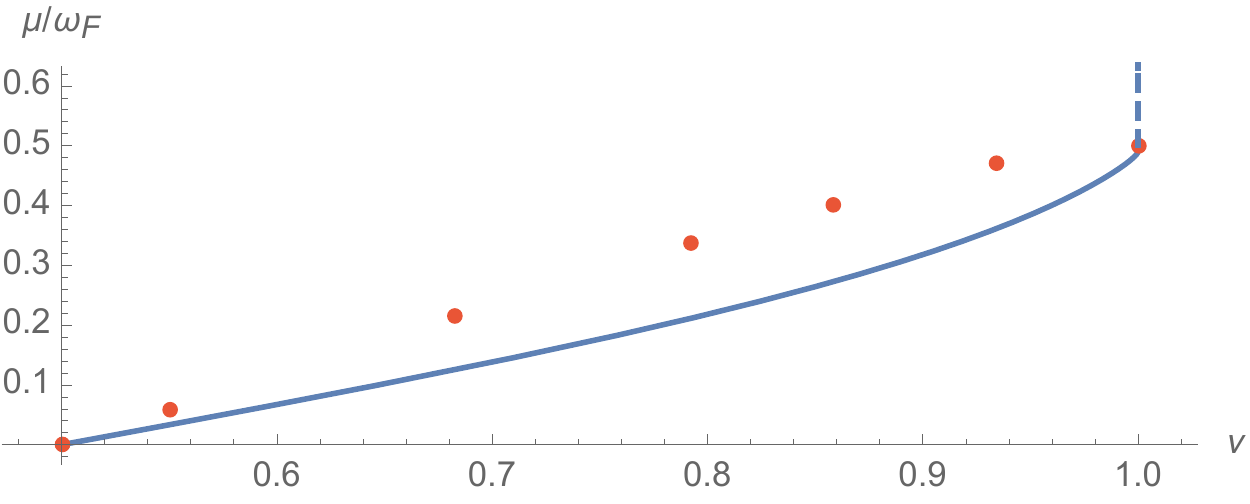}
\caption{
 Same as in Fig. \protect\ref{fig:1}, but for  $N =60M$. In this case
  $x_0 = 0.45$.  }
\label{fig:2}
\end{figure}

\begin{figure}[t]
\includegraphics[width=.9\columnwidth]{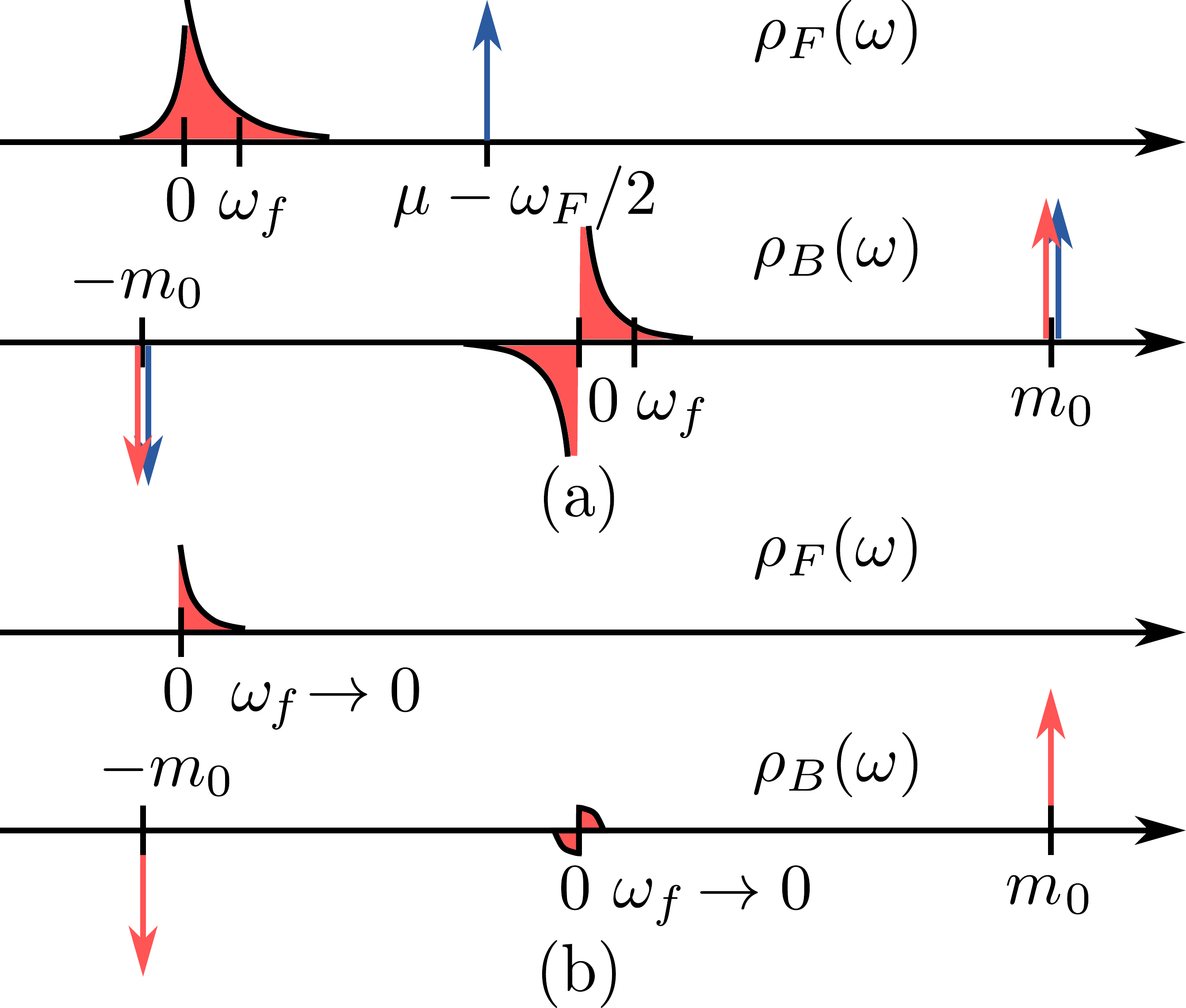}
\caption{Schematic bosonic and fermionic spectral functions,
$\rho_F(\omega)\equiv -\Imm G(\omega+i\eta)$ and $\rho_B(\omega)\equiv\Imm D(\omega+i\eta)$, at the first order transition
for $N\sim M$ (Panel (a)), and at the second-order transition
for
$N/M \to \infty$ (Panel (b)).
The vertical arrows represent $\delta$-function peaks.
 The blue and red peaks in panel (a)
 come from the
 coexisting phases: the NFL phase (red) and the incompressible phase (blue).
}
\label{fig:spec}
\end{figure}

At $N\gg M$ the range, where
 $\mu (\nu)$
   is non-monotonic, shrinks
   (this follows from
   Eq.~\eqref{eq:24a}),    and
    the transition becomes    weakly first-order.  We show this behavior in Fig.~\ref{fig:2} for $N/M =60$. We expect that at $N/M \to \infty$ the transition becomes second order at $\nu =1$ (and $\mu = \omega_F/2$, $x=1/2$, $\alpha =1$).
 In this case, there exists a quantum-critical point that
  separates  a gapless NFL phase (which is by itself quantum critical) and an incompressible, insulating phase. This
 transition has an unconventional feature on its own:  the
 peak in  the bosonic $D(\Omega)$ at $\Omega =
 m_0$ is present on both sides of the transition. In addition,
  a nonzero bosonic spectral weight builds up at small frequencies in the NFL phase and progressively takes
  the spectral weight from the peak at $m_0$.
     The behavior of the  fermionic spectral function is more conventional: the
    gap in the fermionic spectral function $\mu - \omega_F/2$ vanishes at the transition and
     an asymmetric spectral weight builds in  the NFL phase. We  show the behavior of the spectral functions in Fig.~\ref{fig:spec}(b) and present more details in Appendix \ref{sec:end}.

\section{Summary}~~~
In this communication we analyzed the behavior of $M$
 flavors of fermions, randomly interacting with $N$  flavors of massive bosons (the Y-SYK model) away from half-filling.  We showed that the system can be in one of  the  two phases --- a NFL phase with asymmetrically broadened spectral weight, and an incompressible gapped phase. We studied the quantum phase transition between these two phases upon the variation of fermionic density.
  We showed by analytical and numerical calculations that the transition is in general first-order, but becomes second-order in the limit $N/M\to \infty$.
     In the case of the first-order transition, there is a gap filling behavior in the transition region in both fermionic and bosonic sectors. For the second order transition, fermionic gap closes at the transition, but the bosonic spectral function still displays a gap filling behavior.

      Recent numerical studies~\cite{azeyanagi-2018,patel-sachdev-2019} of the complex SYK model also indicated that the system undergoes a  first-order transition upon varying the density.  In distinction to our analysis, there the NFL exponent $x$ is fixed, while in our case it varies with the filling. The possibility of second-order transition at $N/M \to \infty$ was not addressed in these numerical studies.

We conclude by listing several open questions.  First, in our analysis we focused on the case $T=0$. It is possible that at a finite $T$ the first-order transition extends to a line, which terminates at a classical critical point, like in a water-vapor phase diagram. Second,  we analyzed the two-point Green's functions.  It will be interesting to examine the behavior of four-point functions, possibly using the conformal reparametrization symmetry of the low-energy theory. This will shed light on the issue of the strength of superconducting and charge fluctuations.
  Third,  we focused on the
   weak coupling    case,
    $\omega_0\ll m_0$. At  strong coupling,
   the  analysis becomes more involved,
      even though the large $M,N$ limit still guarantees the validity of the self-consistent Schwinger-Dyson equations. It has been pointed out that purely
        bosonic SYK-like models may exhibit glassy behavior at low temperatures~\cite{baldwin-swingle-2019, tulipman-berg-2020}. It would be interesting to see if that happens at strong coupling in Y-SYK model. Finally, in terms of measurable quantities, e.g., in transport experiments, we note that recently it has been proposed that the thermoelectric power~\cite{kruchkov-2020}, measured in quantum matter systems, can be a direct probe of the Bekenstein-Hawking entropy of the SYK models. In SYK-like models it has also been argued that the soft reparametrization modes lead to a universal linear-in-temperature resistivity~\cite{guo-gu-sachdev-2020}. We also note that these transport quantities can be studied using determinant quantum Monte Carlo methods, since the Y-SYK model is free from the fermion sign problem~\cite{pan-2020}.  
We leave the analysis of the transport properties in the Y-SYK model to future work.

 \acknowledgments{We thank
  L.~Classen,   I.~Esterlis, Z.~Meng,   A.~Kamenev,
  S.~Kivelson, S.~Sachdev, J.~Schmalian, P.~Zhang, and especially Y.~Gu for illuminating discussions.  This research was initiated at Stanford University through the Gordon and Betty Moore Foundation's EPiQS Initiative through Grant GBMF4302 and GBMF8686, and at the Aspen Center for Physics, supported by NSF PHY-1066293. AVC is supported by the NSF DMR-1834856. YW is supported by startup funds at University of Florida.}

\onecolumngrid
 
 \appendix
 \vspace{10mm}

The Appendix is organized as follows. In Sec.~\ref{sec:nfl} we provide details on the power-law solution of the bosonic and fermionic self-energies at low frequencies. In Sec.~\ref{sec:nSC} we discuss the analytical solution of the chemical potential $\mu$ by extending the low-energy power-law expression for the self-energies to all frequencies. Despite not being self-consistent at high-energies, we argue that it captures the correct qualitative behavior. In Sec.~\ref{sec:high} we analyze the behavior of the self-energies at high frequencies.  In Sec.~\ref{sec:lutt} we derive an exact relation between the fermionic filling number $\nu$ and the low energy behavior of the fermionic Green's function. In Sec.~\ref{sec:end} we analyze the behavior of the bosonic and fermionic spectral functions at the endpoint of the NFL state and prove that the relation obtained in Sec.~\ref{sec:nSC} is exact.

\section{Details on the NFL solution}\label{sec:nfl}
For the self-energies at low frequencies, we take the ansatz
\begin{align}
\label{eq:0}
\tilde\Sigma(i\omega) \equiv& \Sigma(i\omega) - \Sigma(0) = \omega_f \left|\frac{\omega}{\omega_f}\right|^x (i\sgn(\omega) + \alpha)\nonumber\\
\tilde\Pi(i\Omega) \equiv& \Pi(i\Omega) - \Pi(0) = \beta m_0^2 \left|\frac\Omega{\omega_f}\right |^{1-2x}
\end{align}
There are four dimensionless parameters to solve for: $\alpha,\beta,x$ and $\omega_f/\omega_F$. There are also four equations:
\begin{align}
\label{eq:1}
\Sigma(0)=&-\mu = -\frac{\omega_0^3}{2\pi}\int
\frac{d\omega }{i\omega + \tilde\Sigma(\omega)} \frac{1}{\tilde\Pi(\omega)},\\
\label{eq:2}
\Pi(0) = & -m_0^2 =
 \frac{2M}{N}
 \frac{\omega_0^3}{2\pi}\int \frac{d\omega}{(i\omega + {\tilde \Sigma}(\omega))^2}, \\
\label{eq:3}
\beta m_0^2 \left|\frac{\Omega}{\omega_f}\right|^{1-2x} = &
\frac{M \omega_0^3}{N\pi}\int
\[
{\tilde \Sigma}^{-1}(i\omega){\tilde \Sigma}^{-1}(i\omega+i\Omega) - \tilde\Sigma^{-2}(i\omega)\] d\omega \\
\label{eq:4}
\(i\sgn(\omega)+\alpha \)\omega_f\left|\frac{\omega}{\omega_f}\right|^x = & -\frac{\omega_0^3}{2\pi} \int \[\tilde \Sigma^{-1}(i\Omega) \tilde\Pi^{-1}(i\Omega-i\omega) - \tilde \Sigma^{-1}(i\Omega) \tilde\Pi^{-1}(i\Omega)\]d\Omega
\end{align}

In this section we self-consistently solve Eqs.~(\ref{eq:3}, \ref{eq:4}). Since the integrals are fully convergent using \eqref{eq:0}, the relations we obtain are independent of the high-energy non-universal details.
From Eq.~\eqref{eq:3} we have
\begin{align}
&\beta m_0^2 \left|\frac{\Omega}{\omega_f}\right|^{1-2x} =
\frac{2 M\omega_0^3}{N \omega_f^2}\int \[ \frac{(-i\sgn(\omega)+\alpha)(-i\sgn(\omega+\Omega)+\alpha)\omega_f^{2x}}{(1+\alpha^2)^2|\omega|^{x}|\omega+\Omega|^{x}}-\frac{(\alpha^2-1) \omega_f^{2x}}{(1+\alpha^2)^2|\omega|^{2x}} \]\frac{d\omega}{2\pi},
\end{align}
from which we obtain
\begin{align}
1=-
\frac{M \omega_F}{N \pi\beta\omega_f(1+\alpha^2)^2}\[(1-\alpha^2)\frac{\Gamma^2(-x)}{2\Gamma(-2x)}\frac{(1+\sec\pi x)}{1/x-2}+2\alpha^2\frac{\Gamma^2(-x)}{2\Gamma(-2x)}\frac{1}{1/x-2} \]
\label{eq:21}
\end{align}
From Eq.~\eqref{eq:4}  we have
\begin{align}
\omega_f\[i\sgn(\omega)+\alpha\]\left|{\omega}\right|^x = &- \frac{\omega_0^3}{2\pi\beta m_0^2} \int \[\frac{\alpha-i\sgn(\Omega)}{\alpha^2+1} \left|{\Omega}\right|^{-x} |\omega-\Omega|^{2x-1} - \frac{\alpha}{\alpha^2+1}|\Omega|^{x-1}\]d\Omega.
\end{align}
The real and imaginary parts actually yield a single equation: Indeed, for real and imaginary parts we get respectively
\begin{align}
1=&-\frac{\omega_F}{2\pi\beta\omega_f(1+\alpha^2)}\int \(\frac{1}{|y|^{x}|1-y|^{1-2x}} -\frac{1}{|y|^{1-x}}\)dy\nonumber\\
1=&\frac{\omega_F}{2\pi\beta\omega_f(1+\alpha^2)} \int \frac{\sgn(y)dy}{|y|^x|1-y|^{1-2x}}.
\label{eq:49}
\end{align}
It is straightforward to verify that they lead to a single constraint:
\be
1= -\frac{\omega_F}{4\pi\beta\omega_f(1+\alpha^2)}\frac{\Gamma^2(-x)}{\Gamma(-2x)}.
\label{eq:24}
\ee
Combining \eqref{eq:21} and \eqref{eq:24} we get
\be
\(1-\alpha^2\)\frac{(1+\sec \pi x)}{1/x-2} +\frac{2\alpha^2}{1/x-2} = \frac{N}{2M} \left(1+\alpha^2\right),
\label{eq:24a}
\ee
Eqs.~(\ref{eq:24}, \ref{eq:24a}) are Eqs.~(6, 7) in the main text.
\\

\noindent{{\fbox{
  \parbox{.97\textwidth}{
In evaluating the integrals above we have made use of
\begin{align}
\int \frac{\sgn(y)dy}{|y|^x|1-y|^{1-2x}} =& -\frac{\Gamma^2(-x)}{2\Gamma(-2x)}\\
\int dy\[\frac{ \sgn(y+1/2)\,\sgn(y-1/2)}{|y+1/2|^x\,|y-1/2|^x} - \frac{1}{|y|^{2x}}\]=& \frac{\Gamma^2(-x)}{2\Gamma(-2x)}\, \frac{(1+\sec{\pi x})}{1/x-2}  \\
\int_{-1/2}^{1/2} \frac{dy}{|y+1/2|^x\,|y-1/2|^x}=& -\frac{\Gamma^2(-x)}{2\Gamma(-2x)}\, \frac{1}{1/x-2}
\label{int}
\end{align}

}}}}\\

\subsection{The other two equations on the parameters}
%Calculation of $\mu$ using power-law self-energies}
\label{sec:nSC}

We now consider the other two equations, Eqs.~(\ref{eq:1}, \ref{eq:2}).  One can easily verify that
 relevant frequencies in the integrals over $\omega$ in the r.h.s. of these two equations are of order $\omega_f$.
  At such frequencies, the corrections to the power-law forms in Eq.~(\ref{eq:0}) are of order one. To find the exact values of the integrals in (\ref{eq:1}) and (\ref{eq:2}) one then needs to know the full forms of
   $\Sigma (i\omega)$ and $\Phi (i\omega)$. We found the full forms numerically and combined the numerical results with the two exact relations, Eqs. (\ref{eq:24}) and (\ref{eq:24a}) to obtain the expressions for $x$, $\alpha$, $\beta$, and $\omega_f/\omega_F$ as functions of $\mu$. We then used the additional exact relation between the filling $\mu$ and $\alpha$ and $x$ and expressed $\mu$ as a function of $\nu$ (the dotted lines in Figs.~1 and 2 in the main text.

   Here, we obtain the approximate analytical relations from Eqs.~(\ref{eq:1}, \ref{eq:2}) by keeping the power-law forms of the bosonic and fermionic self-energies. The integrals on the right hand side of (\ref{eq:1}) and (\ref{eq:2}) are ultra-violet convergent because of the bare $\omega$ in the fermionic propagator. Evaluating the integrals, we obtain
   \beq
    \beta = \frac{1}{4^{x-1}} \frac{\Psi_3 (x, \alpha)}{\Psi_1 (x, \alpha)},~~
   \alpha = \frac{\mu}{\omega_F} \frac{4^x}{8} \frac{\Psi_3 (x, \alpha)}{\Psi_1 (x,\alpha) \Psi_2 (x,\alpha)}
\label{ch_1}
\eeq
where
\bea
\Psi_1 (x, \alpha) &=& \frac{M}{N \pi} \int_0^\infty \frac{dy}{y^{2x}} \frac{(1+y^{1-x})^2 -\alpha^2}{((1+y^{1-x})^2 +\alpha^2)^2} \nonumber \\
\Psi_2 (x, \alpha) &=& \frac{1}{2\pi} \int_0^\infty \frac{dy}{y^{1-x}} \frac{1}
{(1+z^{1-x})^2 +\alpha^2} \nonumber\\
\Psi_3 (x,\alpha) &=& = - \frac{M}{2N \pi} \frac{1}{4^x (1+ \alpha^2)} \frac{\Gamma^2(-x)}{\Gamma(-2x)}
\label{ch_2}
\eea
In the last line in (\ref{ch_2}) we used the exact relation (\ref{eq:24a}).
 This relation also allows one to express the spectral asymmetry parameter $\alpha$ via the exponent $x$ (or vise versa), hence $\Psi_{1,2,3}$ are in fact the functions of only one variable.  As a consequence, we can write
    \be
\beta = F_1 (x),~~ \alpha = \frac{\mu}{\omega_F}  F_2(x)
\label{eq:mu}.
\ee
where $F_1 (x) =  4^{x-1} \Psi_3 (x, \alpha (x)) /\Psi_1 (x, \alpha (x))$ and 
$F_2 (x) = 2^{2x-3} \Psi_3 (x, \alpha (x))/(\Psi_1 (x,\alpha (x)) \Psi_2 (x,\alpha (x))$.

 We will see in Sec.~\ref{sec:lutt} that $x$ gradually increases with increasing filling $\nu$ and reaches $x=1/2$ at maximum possible $\nu =1$. Eq.~(\ref{eq:24a}) shows that for $x \to 1/2$, $\alpha \to 1$, and
  $\alpha -1 \approx \pi (1/2-x)$. Substituting this into Eq.~(\ref{ch_2}), we obtain that $\Psi_1 (x, \alpha)$ and $\Psi_3 (x, \alpha)$ vanish in this limit, but the ratio  $\Psi_1 (x, \alpha)/\Psi_3 (x, \alpha)$ tends to $1/2$.  The vanishing of $\Psi_1$ implies that $\omega_f (x) = 2\omega_F \Psi_1 (x, \alpha (x))$ vanishes, i.e., for $x \to 1/2$,  the width of the NFL region shrinks to zero.  Simultaneously, $\Psi_2 (1/2,1) = 1/4$, hence $\mu = \omega_F/2$,  the same as the boundary of the incompressible phase. Using Eqs.~(\ref{ch_1}, \ref{ch_2}, \ref{eq:24}) and the properties of $\Psi_{1,2,3}$, it is straightforward to verify that the functions $F_{1,2} (x)$ are regular $\mathcal{O}(1)$ functions of $x$, with
\be
F_1(1/2) = 1, ~~F_2 (1/2) =1/2.
\ee
Remarkably, both of these special values are exact, as we will show in Sec.~\ref{sec:end}.

\begin{figure}
\includegraphics[width=0.49\columnwidth]{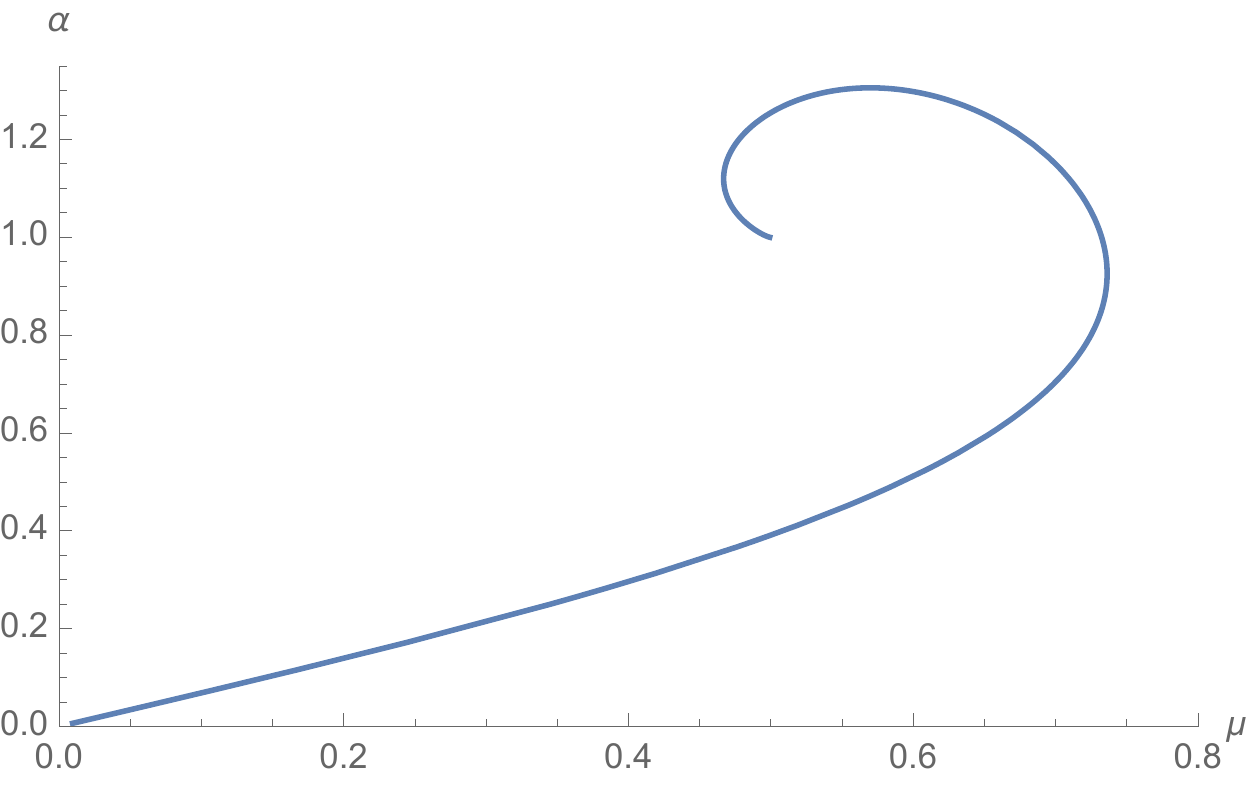}
\includegraphics[width=0.49\columnwidth]{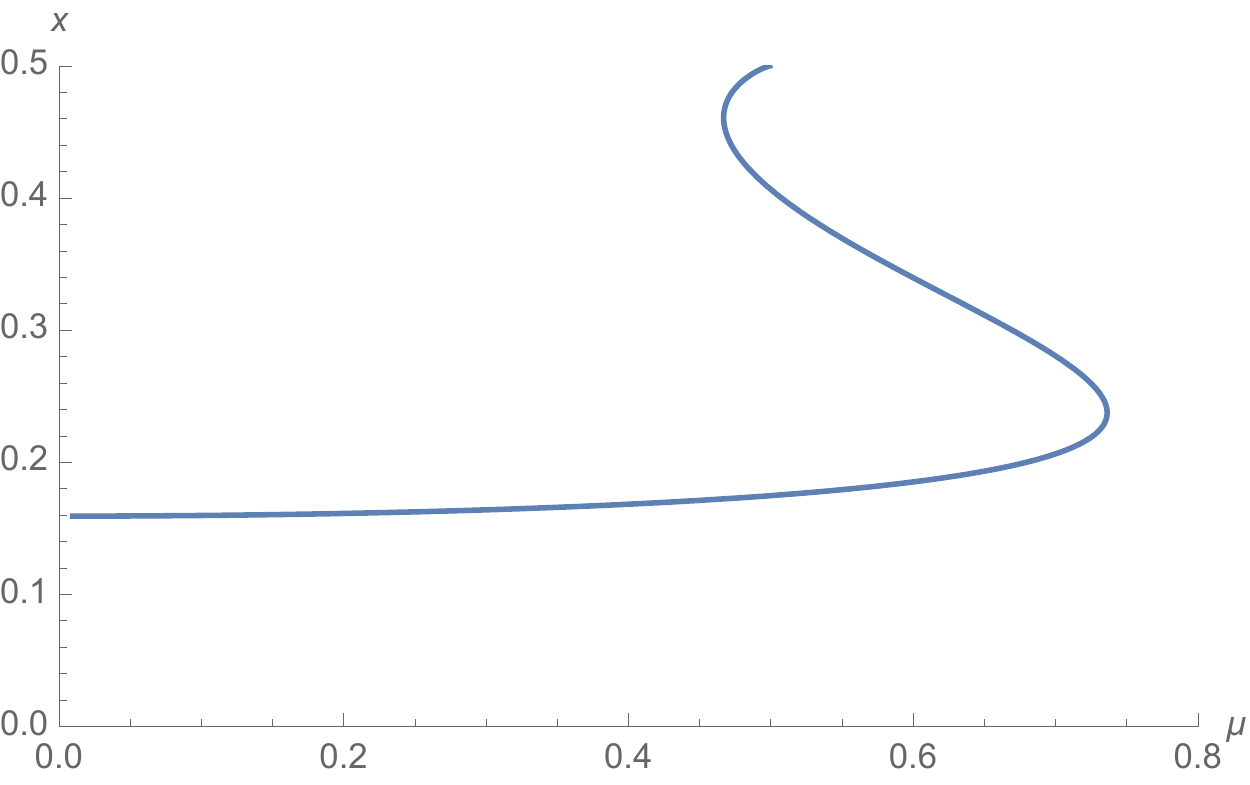}
\caption{The functions $\alpha(\mu)$ and $x(\mu)$ for $M=N$ obtained in Sec.~\ref{sec:nSC}. }
\label{fig:func}
\end{figure}

 In Fig.~\ref{fig:func} we plot $\alpha$ and $x$ as functions of $\mu$ for $M=N$. We see that both become multi-valued functions in some range of $\mu$.  This is a clear indication that the transition between the NFL phase and the incompressible phase may be first order, although to fully address this issue we need to know the relation between the filling $\nu$ and the chemical potential $\mu$.

\subsection{Behavior of the NFL
 solution
  at high energies}\label{sec:high}
In this section we analyze the behavior of the fermionic and bosonic self-energies in the NFL phase at frequencies larger than the  $\omega_f$. For simplicity here we focus on the case well inside the NFL phase, where $\omega_f$ and $\omega_F\equiv \omega_0^3/m_0^2$ are of the same order, which we will use interchangeably for a qualitative analysis.

At $\omega\gg \omega_f$, the fermionic self-energy receives contributions from typical bosonic and fermionic frequencies smaller and greater than $\omega_f$.  We assume and verify that above the NFL energy scale, $\omega, \Omega\gg \omega_f$, both the fermionic and bosonic Green's functions take their free forms.  We then have
\begin{align}
\Sigma(i\omega) \sim& -\frac{\omega_0^3}{m_0^2}\int_{-\omega_f}^{\omega_f} \frac{d\Omega}{2\pi}\frac{1}{|\Omega/\omega_f|^{1-2x}}\frac{1}{i\omega} -\frac{\omega_0^3}{\omega^2+m_0^2} \int_{-\omega_f}^{\omega_f}\frac{d\omega'}{2\pi}\frac{1}{(i\sgn(\omega')-\alpha)\omega_f|\omega'/\omega_f|^x } \nonumber\\
&-{\omega_0^3}\int_{\{\Omega_0\}}\frac{d\Omega}{2\pi}\frac{1}{\Omega^2+m_0^2}\frac{1}{i(\omega-\Omega)},
\end{align}
where the integration domain $\{\Omega_0\}$ excludes $\{-\omega_f,\omega_f\}\cap\{\omega-\omega_f,\omega+\omega_f\}$.
 We have also dropped the chemical potential terms in the bare Green's function, since in the NFL phase $\mu\lesssim \omega_f$ and is subdominant at high frequencies.

The first two integrals are elementary. The third term, in the $\omega_f\ll m_0$ limit can be replaced with a principal value integral circumventing $\omega=\Omega$. We have
\begin{align}
\Sigma(i\omega) =& ic_1\frac{\omega_f^2}{\omega} +c_2\frac{\omega_0^3}{\omega^2+m_0^2} \frac{\alpha}{1+\alpha^2} -{\omega_0^3}\int_{-\infty}^{\infty}\frac{d\Omega}{2\pi}\frac{1}{\Omega^2+m_0^2}\mathrm{P}\[\frac{1}{i(\omega-\Omega)}\]\nonumber\\
=&ic_1\frac{\omega_f^2}{\omega} +c_2\frac{\omega_0^3}{\omega^2+m_0^2} \frac{\alpha}{1+\alpha^2} + i\frac{\omega_0^3}{m_0}\frac{\omega}{\omega^2+m_0^2}
\end{align}
For $\omega_f\ll \omega \ll \sqrt{\omega_0^3/m_0}$, its asymptotic behavior is given by
\be
\Sigma(i\omega) = ic_1\frac{\omega_f^2}{\omega} + c_2\frac{\alpha\omega_f}{1+\alpha^2} =  ic_1\frac{\omega_f^2}{\omega} + \tilde c_2\mu,
\ee
where in the last step we have used $\mu\sim \alpha\omega_f$. Combined with the result on $\Sigma(i\omega)$ at $\omega\ll \omega_f$, we see that the two asymptotic behaviors do match at $\omega\sim \omega_f$.

For $\omega\gg\sqrt{\omega_0^3/m_0}$, the self-energy has a non-monotonic behavior, given by
\be
\Sigma(i\omega) =c_2\frac{\omega_0^3}{\omega^2+m_0^2} \frac{\alpha}{1+\alpha^2} + i\frac{\omega_0^3}{m_0}\frac{\omega}{\omega^2+m_0^2},
\ee
in which $\Imm (\Sigma(i\omega))$ remains a constant $\sim \mu$ at $\omega\ll m_0$ and decays to zero at $\omega\gg m_0$. On the other hand $\Re (\Sigma(i\omega))$ first increases linearly and then again decreases as $1/\omega$.
However, this non-monotonic behavior in has little effect on the low-energy behaviors of the system. For both regimes above, the self-energy effects in the fermionic Green's function is small, since $\omega \ll |\Sigma(i\omega)|$, thus justifying our assumption that the fermions are essentially free.

For the bosons at $\Omega\gg \omega_f$, the self energy comes from frequencies $\omega\gg\omega_f$. We have qualitatively
\be
\Pi(i\Omega) \sim -\int_{\Omega}\frac{\omega_0^3 d\omega}{\omega^2} \sim \frac{\omega_0^3}{\Omega} \ll m_0^2,~~ \tilde\Pi(i\Omega)\approx m_0^2.
\ee
 We see that
  the bosons are essentially free in this frequency range.

\begin{figure}
\includegraphics[width=0.49\columnwidth]{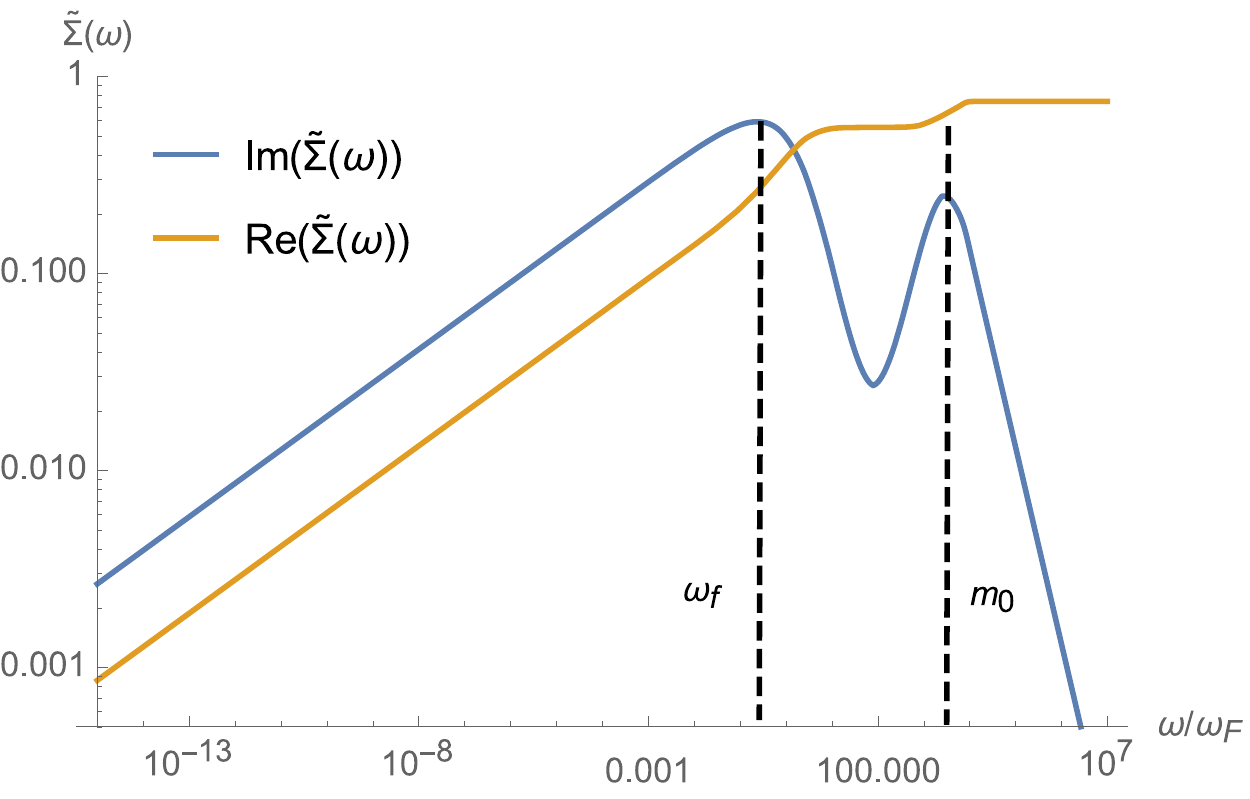}
\includegraphics[width=0.49\columnwidth]{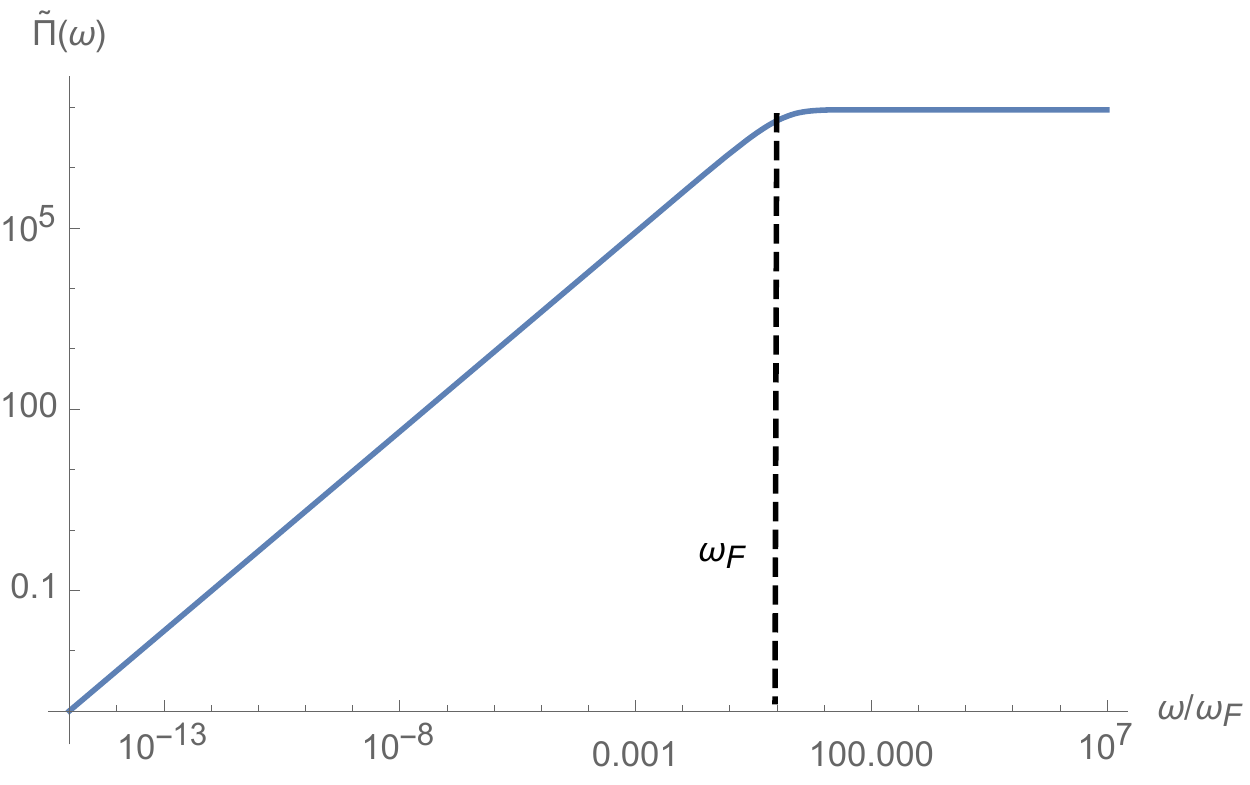}
\caption{The full solution $\tilde\Sigma(i\omega)$ and $\tilde\Pi(i\omega)$ for the  Schwinger-Dyson equations. We have taken $m_0=3000\omega_F$.}
\label{fig:s1}
\end{figure}

These results can be directly verified by numerically solving the Schwinger-Dyson equations, using the same iterative technique sketched in the main text. In Fig.~\ref{fig:s1} we present the numerical solution for $\tilde\Sigma(i\omega)$ and $\tilde\Pi(i\omega)$, in which the power-law behavior in different regimes can be easily seen. The $\mathcal{O}(1)$ numerical constants $c_1$ and $c_2$ can be fitted from the numerics.

\section{Luttinger's theorem for the NFL state}\label{sec:lutt}
In this section we focus on the analytical derivation of the filling fraction $\nu$. From the usual relation
\be
\nu= \int_{-i\infty}^{i\infty}\frac{dz}{2\pi i}G(z)e^{z0_+},
\ee
where $G(z)=1/(z-\tilde\Sigma(z))$,
 it may seem that its value depends on details of the system behavior at high energies, but due to a Luttinger-theorem-like relation it can be shown to be completely determined by low-energy, or infrared (IR) behaviors. This is an example of ultraviolet(UV)-IR mixing, a feature shared in the original complex SYK model, and its mathematical structure is similar to the chiral anomaly in quantum electrodynamics. Our presentation closely parallels that in Ref.~\onlinecite{gu-2020}.

Using the identity $\partial_z(G^{-1}(z)-\tilde\Sigma(z))=1$, we rewrite $\nu$ as
\be
\nu= \int_{-i\infty}^{i\infty}\frac{dz}{2\pi i}G(z)e^{z0_+}  =\int_{-i\infty}^{i\infty}\frac{dz}{2\pi i} G(z) e^{z0_+} \partial_z(G^{-1}(z)-\tilde\Sigma) \equiv I_1-I_2
\label{eq:nu2}
\ee
The first term $I_1$ is expressed as
\begin{align}
I_1 =& \int_{-i\infty}^{i\infty}\frac{dz}{2\pi i} G(z) e^{z\delta} \partial_z G^{-1}(z) \nonumber\\
=& \int_{-i\infty}^{i\infty}\frac{d \log (G^{-1}(z))}{2\pi i} e^{z\delta}
\end{align}
This integral is IR divergent. But it is clear that in the first equation of \eqref{eq:nu2} there is no IR divergence, which is convergent at $z=0$. Thus the IR divergence should cancel when properly regularized. One way to properly regularize  both integrals  is to take principle values. For $I_1$,
\begin{align}
I_1 =& -\mathrm{P}\int_{-i\infty}^{i\infty}\frac{d \log (G^{-1}(z))}{2\pi i} e^{z\delta} =  \(\int_{-i\eta}^{-\infty-i\eta}-\int_{i\eta}^{-\infty+i\eta}\) \frac{d \log (G^{-1}(z))}{2\pi i}\nonumber\\
=&\frac{\arg(G^{-1}(-i\eta))-\arg(G^{-1}(-\infty+i\eta))}{\pi} = \frac{\arg(G^{-1}(-i\eta))}{\pi}-1
\label{eq:I1}
\end{align}
where in the second step we have deformed the integration contour allowed by the $e^{z\delta}$ factor.
Interestingly this only depends on IR properties of the Green's functions.

The evaluation of $I_2$ is more tricky. If we simply use IR expressions for $G$ and $\tilde\Sigma$ in,
\be
I_2 = \int_{-i\infty}^{i\infty}\frac{dz}{2\pi i}G(z) \partial_z\tilde \Sigma(z)e^{z\delta}
\ee
 the $I_2$ integral is formally logarithmically divergent. However, the infinities at both the IR and UV ends are canceled by the integration domain (the asymmetry cancels out between $G$ and $\Sigma$). All we need is to properly regularize the integral at both ends.

 In the standard treatment of the Luttinger's theorem~\cite{luttinger-1960,altshuler-1998}, the $I_2$ term can be transformed to an integral over a total derivative (of the Luttinger-Ward functional) and thus is zero. In our case however, the integral is not along a continuous path, since we are taking principal values. Across the branch cuts of $G(z)$ there are additional boundary terms, which are divergent. They can be made to exactly cancel each other if we take $e^{z\delta}\to 1$ prior to the principal value limit, but physically from the definition of the charge one should take the $e^{z\delta}\to 1$ limit only \emph{after} the principal value limit. In fact, according to the Riemann rearrangement theorem they can be made to take any value, so the correct order of limits is crucial. From this analysis, we see that the nonzero contribution should come from the IR end, since the boundary terms at infinity is well-behaved and vanishes.

 We deform the integration contour in the same way as in $I_1$, and
 \be
 I_2= \int_{0}^{-\infty} \frac{d\omega}{\pi} \Imm \[G(iz )\partial_{z} \Sigma(iz)\]\big|_{iz\to \omega+i\eta} e^{x\delta}
 \ee
 It is important to not directly replace the integral $\partial_{iz}$ with $\partial_\omega$ since the self-energy is non-analytic along the real axis.
 This integral is well behaved, so we can take the $\delta \to 0$ limit first. As we will see, the integral converges even if we just use the power-law form for the self-energy, since the UV logarithmical divergence cancels when the integration domain is folded to a half space.

For convenience, let us rewrite our self-energies at small frequencies as
 \begin{align}
 \tilde \Sigma(\pm i\omega) &=\pm i e^{\mp i\theta}  \bar\omega_f^{1-x} \omega^x\nonumber \\
 \tilde \Pi(\pm i\Omega) &= \bar\beta m_0^2 \(\frac{\Omega}{\bar\omega_f}\)^{1-2x}
 \end{align}
 where we have defined $\theta=\tan^{-1} \alpha$ and $\bar\beta \bar \omega_f= (1+\alpha^2) \beta \omega_f$.

 The $I_1$ integral in Eq.~\eqref{eq:I1} yields
 \be
 I_1 = \frac{1}{2} +\frac{\theta}{\pi}.
 \ee
 It turns out that it is easiest to analyze the $I_2$ integral using the spectral representation, which gives
 \begin{align}
 \frac{I_2}{\omega_0^3} =& \int_{-\infty}^{0}\frac{d\omega}{\pi} \Imm\[\int \frac{d\omega_1}{\pi} \frac{\rho_F(\omega_1)}{iz-\omega_1}\int\frac{d\Omega}{2\pi} \int \frac{d\omega_2d \omega_3}{\pi^2} \frac{-\rho_F(\omega_2)}{(iz-i\Omega-\omega_2)^2}\frac{\rho_B(\omega_3)}{i\Omega-\omega_3}\]\bigg|_{iz\to \omega+i\eta} \nonumber\\
=&\int_{-\infty}^{0}\frac{d\omega}{\pi} \Imm\[\int \frac{d\omega^3}{\pi^3} \frac{\rho_F(\omega_1)}{\omega-\omega_1+i\eta}\frac{\rho_F(\omega_2)\rho_B(\omega_3)}{(\omega-\omega_2-\omega_3+i\eta)^2}(\sgn(\omega_2)+\sgn(\omega_3))\]\nonumber\\
=&-\int^{\infty}_{0}\frac{d\omega}{\pi} \Imm\[\int \frac{d\omega^3}{\pi^3} \frac{\rho_F(-\omega_1)}{-\omega+\omega_1-i\eta}\ \frac{\rho_F(-\omega_2)\rho_B(-\omega_3)}{(-\omega+\omega_2+\omega_3-i\eta)^2}(-\sgn(\omega_2)-\sgn(\omega_3))\]\nonumber\\
=&-\int^{\infty}_{0}\frac{d\omega}{\pi} \Imm\[\int \frac{d\omega^3}{\pi^3} \frac{\rho_F(-\omega_1)}{\omega-\omega_1+i\eta}\ \frac{\rho_F(-\omega_2)\rho_B(-\omega_3)}{(\omega-\omega_2-\omega_3+i\eta)^2}(\sgn(\omega_2)+\sgn(\omega_3))\]\nonumber\\
=&\int^{\infty}_{0}\frac{d\omega}{\pi} \Imm\[\int \frac{d\omega^3}{\pi^3} \frac{\rho_F(-\omega_1)}{\omega-\omega_1+i\eta}\ \frac{\rho_F(-\omega_2)\rho_B(\omega_3)}{(\omega-\omega_2-\omega_3+i\eta)^2}(\sgn(\omega_2)+\sgn(\omega_3))\]
\label{eq:I2}
\end{align}
Here the spectral functions are given by $\rho_F(\omega)= - \Imm G(\omega+i\eta)$ and $\rho_B(\omega)=  \Imm D(\omega+i\eta)$.
In our case we have for $0<\omega\ll \omega_f$,
\begin{align}
\rho_F(\pm\omega)&=-\Imm\[ G( \pm \omega+ i\eta)\] =\Imm\[ i e^{i\theta}  \bar\omega_f^{x-1} (\mp i\omega)^{-x}\] = \cos \(\frac{\pi x}{2}\pm\theta\)\bar\omega_f^{x-1}\omega^{-x}\nonumber \\
\rho_B(\pm\omega)&=\Imm\[ D(\pm\omega + i\eta)\] = \Imm\[\frac1{\bar\beta m_0^2} \(\frac{\mp i\omega}{\bar\omega_f}\)^{2x-1}\] = \pm \frac{\cos(\pi x)}{\bar\beta m_0^2} \(\frac{\omega}{\bar\omega_f}\)^{2x-1}
\label{eq:spec}
\end{align}
 In the last step of Eq.~\eqref{eq:I2} we have used the fact $\rho_B(-\omega)=-\rho_B(\omega)$. On the other hand, $\rho_F(\omega)$ is asymmetric.

Averaging the second and the fifth lines of Eq.~\eqref{eq:I2}, we get by separating the $\rho_F(\omega_1)\rho_F(\omega_2)$ into symmetric and antisymmetric parts
\begin{align}
\!\!\!\! \!\!\frac{I_2}{\omega_0^3} =&\int^{\infty}_{-\infty}\frac{d\omega}{4\pi} \Imm\[\int \frac{d\omega^3}{\pi^3} \frac{\[\rho_F(\omega_1)\rho_F(\omega_2)+\rho_F(-\omega_1)\rho_F(-\omega_2)\]\rho_B(\omega_3)}{(\omega-\omega_1+i\eta)(\omega-\omega_2-\omega_3+i\eta)^2}(\sgn(\omega_2)+\sgn(\omega_3))\] \nonumber \\
+&\int^{\infty}_{0}\frac{d\omega}{2\pi} \Imm\[\int \frac{d\omega^3}{\pi^3} \frac{\[\rho_F(\omega_1)\rho_F(\omega_2)-\rho_F(-\omega_1)\rho_F(-\omega_2)\]\rho_B(\omega_3)}{(\omega-\omega_1+i\eta)(\omega-\omega_2-\omega_3+i\eta)^2}(\sgn(\omega_2)+\sgn(\omega_3))\]
\end{align}
The contour integral over $x$ in the first line vanishes since all poles are on the same side. From Eq.~\eqref{eq:spec} the second line vanishes unless $\omega_1$ and $\omega_2$ have the same sign. We have
\begin{align}
 {I_2} =&- \frac{\omega_F}{\bar\beta\bar \omega_f}\int_0^\infty  \frac{d\omega}{\pi}\Imm\int_0^\infty\frac{d\omega^3}{4\pi^3}\frac{\sin(2\pi x)\sin(2\theta)}{\omega_1^x\omega_2^x\omega_3^{1-2x}} \frac{1}{(\omega-\omega_1+i\eta)(\omega-\omega_2-\omega_3+i\eta)^2} \nonumber \\
= & -\frac{x \sin(2\theta)}{\sin(\pi x)} \lim_{\eta\to 0}\int_0^\infty  \frac{d\omega}{\pi}\Imm\frac{1}{\omega+i\eta}  =  -\frac{x \sin(2\theta)}{2\sin(\pi x)}.
\end{align}
One may wonder if the different form of self-energy at high frequencies will contribute to a correction to the result, but we note that the contribution to the above result comes from $\omega\sim \eta$, thus the UV contribution is small $\mathcal{O}(\eta/\omega_0)=0$. Hence this result is \emph{exact}. \\
\\
\noindent{{\fbox{
  \parbox{.97\textwidth}{
  We have used the following identities
\begin{align}
&\bar\beta\bar\omega_f=\beta\omega_f(1+\alpha^2)= -\frac{\omega_F}{2\pi}\frac{\Gamma^2(-x)}{2\Gamma(-2x)}.\\
&\int_0^\infty\frac{d\omega^3}{\omega_1^x\omega_2^x\omega_3^{1-2x}} \frac{1}{(y-\omega_1)(y-\omega_2-\omega_3)^2} = \frac{\pi ^{5/2} 4^x \csc (\pi  x) \csc(2 \pi  x) \Gamma (1-x)}{y \Gamma \left(\frac{1}{2}-x\right)} \\
&-\frac{(8 \pi  \Gamma (-2 x)) \left(2^{2 x-1} \Gamma (1-x)\right)}{x \Gamma (-x)^2 \left(\sqrt{\pi } \Gamma \left(\frac{1}{2}-x\right)\right)}=2.
\label{int2}
\end{align}

}}}}\\

 Combining, we get
 \be
 \nu = \frac{1}{2} + \frac{\theta}{\pi} + \frac{x\sin (2\theta)}{2\sin(\pi x)}=\frac{1}{2} + \frac{\tan^{-1} \alpha}{\pi} + \frac{x}{2\sin(\pi x)}\frac{2\alpha}{1+\alpha^2}.
 \label{eq:lutt}
 \ee
 This is Eq.~(10) of the main text.

 Remarkably, this result is identical to the complex SYK$_q$ model, but here our $x$ is continuously tunable.
 It may not immediately clear why $I_2$ is the same  as the SYK model, which has very different  structures. It turns out $\nu$ can also be directly derived from $I_1$ alone, by doing the calculations and regularization in the time domain, in which case $I_2=0$, as shown in Ref.~\onlinecite{gu-2020}. Since $I_1$ only depends on fermion Green's function, our result should be the same as SYK$_q$ model with the same scaling dimension for the fermions.  The fact that $I_{1,2}$ separately depends on regularization scheme but their sum does not is also reminiscent of features in chiral anomaly.

We note in passing that if we used the power-law form for the self-energy, Eq.~(\ref{eq:0}) and computed
$ \nu= \int_{-i\infty}^{i\infty}\frac{dz}{2\pi i}G(z)e^{z0_+}$, we would obtain instead
\beq
\nu = \frac{1}{2} + \frac{1}{\pi} \frac{\tan^{-1} \alpha}{1-x}
\label{ch_3}
\eeq
This relation is similar, but not identical to the exact one, although both yield
$\nu =1$ for $\alpha \to 1$ and $x \to 1/2$.

We also note that if we combine (\ref{ch_3}) with Eq.~(\ref{ch_2}) and the exact relations (\ref{eq:24}, \ref{eq:24a}), we find that very near $\nu =1$, $\mu$ actually increases with $\nu$: $\mu = (\omega_F/2) (1-3.2 (M/N) \sqrt{1-\nu})$, hence   $\partial \nu/\partial \mu >0$. Taken at the face value, this would imply that the transition between the incompressible and the NFL phase is second order for any $N/M$, and then there is a first-order  transition within in NFL phase.  However, this is likely an artifact of the approximation as we didn't detect this behavior in numerical studies.
%end

\section{The endpoint of the NFL solution}\label{sec:end}
In this section we analyze how the NFL solution vanishes at $x=1/2$, in particular at which chemical potential $\mu$ it does so. In general, unlike the filling $\nu$, the chemical potential cannot be analytically expressed by universal quantities. However we find that it takes a universal value at the end of the NFL solution, namely, as filling $\nu\to 1$. While the NFL phase as a stable state disappears through a first-order phase transition, the saddle-point solution to the Schwinger-Dyson equation exists beyond the first-order phase transition. In the limit where the first-order transition becomes second-order, such an endpoint of the NFL solution becomes a true quantum-critical point.

As was shown in the main text, at the endpoint of the NFL solution, $x\to 1/2$, $\alpha\to 1$, and $\beta\omega_f\to 0$.  We know that the bosonic self-energy $\tilde\Pi(i\Omega)$ at $\Omega\gtrsim \omega_f$ crosses over from $\beta m_0
^2|\Omega/\omega_f|^{1-2x}$ to the bare mass $m_0^2$, which is only possible at $x\to 1/2$ for
\be
\beta =1,~\omega_f\to 0.
\ee
Notice that $\beta=1$ is also obtained in Sec.~\ref{sec:nSC}.
In this limiting case calculating $\Sigma(0)$ from the imaginary-time Green's functions can be difficult since the bosonic Green's function involves a $0^0$ limit. However the spectral functions, whose IR behavior are shown in Eq.~\eqref{eq:spec}, are free from this ambiguity.

The chemical potential is given in terms of spectral functions by
\begin{align}
\mu= -\Sigma(0) =& -\omega_0^3\int\frac{d\Omega}{2\pi}\int\frac{d\omega d\omega'}{\pi^2}\frac{\rho_B(\omega) \rho_F(\omega')}{(i\Omega -\omega)(i\Omega- \omega')} \nonumber\\
=&-\omega_0^3\int \frac{d\omega d\omega'}{2\pi^2} \frac{\rho_B(\omega)\rho_F(\omega')}{\omega-\omega'}\[\sgn(\omega)-\sgn(\omega')\].
\end{align}
At $x\to 1/2$, from Eq.~\eqref{eq:spec} we have at low energies ($0<\omega\ll\bar\omega_f$)
\begin{align}
\rho_F(\pm \omega) =& \cos \(\frac{\pi x}{2}\pm\theta(x)\)\bar\omega_f^{-1/2}\omega^{-1/2} \nonumber\\
\rho_B(\pm \omega)=& \pm \frac{\cos(\pi x)}{\bar\beta m_0^2} \(\frac{\omega}{\bar\omega_f}\)^{2x-1}.
\end{align}
Notice that in the $x\to 1/2$ limit the low-energy bosonic spectral weight gets progressively depleted.
This indeed matches the starting point of the gapped phase, in which bosons are gapped at the bare mass and the chemical potential gets renormalized to zero. In other words, upon entering the NFL phase, the boson mass gap ``fills in" rather than ``closes in".  We have showed the behavior of the spectral functions $\rho_B$
in Fig.~3 of the main text.

 Going back to the NFL side, since all the bosonic spectral weight is depleted at low frequencies, at $x\to1/2$, $\rho_B$ is peaked at high-energies, and in this regime the bosons behave like free ones. We have
\be
\rho_B(\pm \omega) = \pm \pi \delta(\omega - m_0)/(2m_0).
 \ee
Thus
\be
\mu =-\frac{\omega_0^3}{2m_0}\[ \int_{-\infty}^0 \frac{d\omega' }{\pi} \frac{\rho_F(\omega')}{m_0-\omega'} -\int^{\infty}_0 \frac{d\omega' }{\pi} \frac{\rho_F(\omega')}{m_0+\omega'} \] .
\ee
In the weak coupling limit, the width of $|\rho_F(\omega)|$ in frequency is the scale of interaction $\omega_f\ll m_0$. Thus approximately we obtain
\be
\mu= -\frac{\omega_0^3}{2m_0^2}\int_{0}^{\infty} \frac{d\omega }{\pi} \[\rho_F(-\omega) -\rho_F(\omega)\] = \frac{\omega_0^3 }{2m_0^2}(2\nu -1)\big|_{x={1}/{2}}.
\ee
In the last step we have used $\nu=\int_0^\infty d\omega \rho_F(\omega)/\pi$ and $1-\nu=\int^0_{-\infty} d\omega \rho_F(\omega)/\pi$, which can be easily seen by a spectral decomposition of the fermionic Green's function.

We have now converted the calculation of $\mu$ to that of $\nu$, to which one can apply the Luttinger theorem. From Eq.~\eqref{eq:lutt} we have $\nu=1$ at $x=1/2$. Therefore, we conclude as the NFL solution approaches its endpoint, the chemical potential tends to
\be
\mu\(x=\frac{1}{2}\) =\frac{\omega_0^3 }{2m_0^2}  \equiv \frac{\omega_F}{2},
\ee
which matches the starting point of the insulator phase. Interestingly, this result is also captured by using the power-law forms of the self-energies. Indeed, we see that the key ingredient for $\mu=\omega_F/2$ is the mismatch of fermionic and bosonic spectral weights at $x=1/2$, which is also present even if one uses power-law forms of the self-energies. Of course, in the insulating phase the chemical potential can be obtained by the same procedures described here. However, this result does not hold for the strong coupling case, where $\omega_0\gg m_0$.
\bibliography{ysyk-bib}

\begin{thebibliography}{85}%
\makeatletter
\providecommand \@ifxundefined [1]{%
 \@ifx{#1\undefined}
}%
\providecommand \@ifnum [1]{%
 \ifnum #1\expandafter \@firstoftwo
 \else \expandafter \@secondoftwo
 \fi
}%
\providecommand \@ifx [1]{%
 \ifx #1\expandafter \@firstoftwo
 \else \expandafter \@secondoftwo
 \fi
}%
\providecommand \natexlab [1]{#1}%
\providecommand \enquote  [1]{``#1''}%
\providecommand \bibnamefont  [1]{#1}%
\providecommand \bibfnamefont [1]{#1}%
\providecommand \citenamefont [1]{#1}%
\providecommand \href@noop [0]{\@secondoftwo}%
\providecommand \href [0]{\begingroup \@sanitize@url \@href}%
\providecommand \@href[1]{\@@startlink{#1}\@@href}%
\providecommand \@@href[1]{\endgroup#1\@@endlink}%
\providecommand \@sanitize@url [0]{\catcode `\\12\catcode `\$12\catcode
  `\&12\catcode `\#12\catcode `\^12\catcode `\_12\catcode `\%12\relax}%
\providecommand \@@startlink[1]{}%
\providecommand \@@endlink[0]{}%
\providecommand \url  [0]{\begingroup\@sanitize@url \@url }%
\providecommand \@url [1]{\endgroup\@href {#1}{\urlprefix }}%
\providecommand \urlprefix  [0]{URL }%
\providecommand \Eprint [0]{\href }%
\providecommand \doibase [0]{http://dx.doi.org/}%
\providecommand \selectlanguage [0]{\@gobble}%
\providecommand \bibinfo  [0]{\@secondoftwo}%
\providecommand \bibfield  [0]{\@secondoftwo}%
\providecommand \translation [1]{[#1]}%
\providecommand \BibitemOpen [0]{}%
\providecommand \bibitemStop [0]{}%
\providecommand \bibitemNoStop [0]{.\EOS\space}%
\providecommand \EOS [0]{\spacefactor3000\relax}%
\providecommand \BibitemShut  [1]{\csname bibitem#1\endcsname}%
\let\auto@bib@innerbib\@empty
%</preamble>
\bibitem [{\citenamefont {Hussey}\ \emph {et~al.}(1998)\citenamefont {Hussey},
  \citenamefont {Mackenzie}, \citenamefont {Cooper}, \citenamefont {Maeno},
  \citenamefont {Nishizaki},\ and\ \citenamefont
  {Fujita}}]{hussey-mackenzie-1998}%
  \BibitemOpen
  \bibfield  {author} {\bibinfo {author} {\bibfnamefont {N.~E.}\ \bibnamefont
  {Hussey}}, \bibinfo {author} {\bibfnamefont {A.~P.}\ \bibnamefont
  {Mackenzie}}, \bibinfo {author} {\bibfnamefont {J.~R.}\ \bibnamefont
  {Cooper}}, \bibinfo {author} {\bibfnamefont {Y.}~\bibnamefont {Maeno}},
  \bibinfo {author} {\bibfnamefont {S.}~\bibnamefont {Nishizaki}}, \ and\
  \bibinfo {author} {\bibfnamefont {T.}~\bibnamefont {Fujita}},\ }\href
  {\doibase 10.1103/PhysRevB.57.5505} {\bibfield  {journal} {\bibinfo
  {journal} {Phys. Rev. B}\ }\textbf {\bibinfo {volume} {57}},\ \bibinfo
  {pages} {5505} (\bibinfo {year} {1998})}\BibitemShut {NoStop}%
\bibitem [{\citenamefont {Stewart}(2001)}]{stewart-2001}%
  \BibitemOpen
  \bibfield  {author} {\bibinfo {author} {\bibfnamefont {G.~R.}\ \bibnamefont
  {Stewart}},\ }\href {\doibase 10.1103/RevModPhys.73.797} {\bibfield
  {journal} {\bibinfo  {journal} {Rev. Mod. Phys.}\ }\textbf {\bibinfo {volume}
  {73}},\ \bibinfo {pages} {797} (\bibinfo {year} {2001})}\BibitemShut
  {NoStop}%
\bibitem [{\citenamefont {Shibauchi}\ \emph {et~al.}(2014)\citenamefont
  {Shibauchi}, \citenamefont {Carrington},\ and\ \citenamefont
  {Matsuda}}]{shibauchi-carrington-matsuda-2014}%
  \BibitemOpen
  \bibfield  {author} {\bibinfo {author} {\bibfnamefont {T.}~\bibnamefont
  {Shibauchi}}, \bibinfo {author} {\bibfnamefont {A.}~\bibnamefont
  {Carrington}}, \ and\ \bibinfo {author} {\bibfnamefont {Y.}~\bibnamefont
  {Matsuda}},\ }\href {\doibase 10.1146/annurev-conmatphys-031113-133921}
  {\bibfield  {journal} {\bibinfo  {journal} {Annual Review of Condensed Matter
  Physics}\ }\textbf {\bibinfo {volume} {5}},\ \bibinfo {pages} {113} (\bibinfo
  {year} {2014})},\ \Eprint
  {http://arxiv.org/abs/https://doi.org/10.1146/annurev-conmatphys-031113-133921}
  {https://doi.org/10.1146/annurev-conmatphys-031113-133921} \BibitemShut
  {NoStop}%
\bibitem [{\citenamefont {Keimer}\ \emph {et~al.}(2015)\citenamefont {Keimer},
  \citenamefont {Kivelson}, \citenamefont {Norman}, \citenamefont {Uchida},\
  and\ \citenamefont {Zaanen}}]{keimer-kivelson-norman-2015}%
  \BibitemOpen
  \bibfield  {author} {\bibinfo {author} {\bibfnamefont {B.}~\bibnamefont
  {Keimer}}, \bibinfo {author} {\bibfnamefont {S.~A.}\ \bibnamefont
  {Kivelson}}, \bibinfo {author} {\bibfnamefont {M.~R.}\ \bibnamefont
  {Norman}}, \bibinfo {author} {\bibfnamefont {S.}~\bibnamefont {Uchida}}, \
  and\ \bibinfo {author} {\bibfnamefont {J.}~\bibnamefont {Zaanen}},\ }\href
  {https://doi.org/10.1038/nature14165} {\bibfield  {journal} {\bibinfo
  {journal} {Nature}\ }\textbf {\bibinfo {volume} {518}},\ \bibinfo {pages}
  {179} (\bibinfo {year} {2015})}\BibitemShut {NoStop}%
\bibitem [{\citenamefont {Varma}\ \emph {et~al.}(1989)\citenamefont {Varma},
  \citenamefont {Littlewood}, \citenamefont {Schmitt-Rink}, \citenamefont
  {Abrahams},\ and\ \citenamefont {Ruckenstein}}]{varma-littlewood-1989}%
  \BibitemOpen
  \bibfield  {author} {\bibinfo {author} {\bibfnamefont {C.~M.}\ \bibnamefont
  {Varma}}, \bibinfo {author} {\bibfnamefont {P.~B.}\ \bibnamefont
  {Littlewood}}, \bibinfo {author} {\bibfnamefont {S.}~\bibnamefont
  {Schmitt-Rink}}, \bibinfo {author} {\bibfnamefont {E.}~\bibnamefont
  {Abrahams}}, \ and\ \bibinfo {author} {\bibfnamefont {A.~E.}\ \bibnamefont
  {Ruckenstein}},\ }\href {\doibase 10.1103/PhysRevLett.63.1996} {\bibfield
  {journal} {\bibinfo  {journal} {Phys. Rev. Lett.}\ }\textbf {\bibinfo
  {volume} {63}},\ \bibinfo {pages} {1996} (\bibinfo {year}
  {1989})}\BibitemShut {NoStop}%
\bibitem [{\citenamefont {Hertz}(1976)}]{hertz-1976}%
  \BibitemOpen
  \bibfield  {author} {\bibinfo {author} {\bibfnamefont {J.~A.}\ \bibnamefont
  {Hertz}},\ }\href {\doibase 10.1103/PhysRevB.14.1165} {\bibfield  {journal}
  {\bibinfo  {journal} {Phys. Rev. B}\ }\textbf {\bibinfo {volume} {14}},\
  \bibinfo {pages} {1165} (\bibinfo {year} {1976})}\BibitemShut {NoStop}%
\bibitem [{\citenamefont {Millis}(1993)}]{millis-1993}%
  \BibitemOpen
  \bibfield  {author} {\bibinfo {author} {\bibfnamefont {A.~J.}\ \bibnamefont
  {Millis}},\ }\href {\doibase 10.1103/PhysRevB.48.7183} {\bibfield  {journal}
  {\bibinfo  {journal} {Phys. Rev. B}\ }\textbf {\bibinfo {volume} {48}},\
  \bibinfo {pages} {7183} (\bibinfo {year} {1993})}\BibitemShut {NoStop}%
\bibitem [{\citenamefont {Oganesyan}\ \emph {et~al.}(2001)\citenamefont
  {Oganesyan}, \citenamefont {Kivelson},\ and\ \citenamefont
  {Fradkin}}]{oganesyan-kivelson-fradkin-2001}%
  \BibitemOpen
  \bibfield  {author} {\bibinfo {author} {\bibfnamefont {V.}~\bibnamefont
  {Oganesyan}}, \bibinfo {author} {\bibfnamefont {S.~A.}\ \bibnamefont
  {Kivelson}}, \ and\ \bibinfo {author} {\bibfnamefont {E.}~\bibnamefont
  {Fradkin}},\ }\href {\doibase 10.1103/PhysRevB.64.195109} {\bibfield
  {journal} {\bibinfo  {journal} {Phys. Rev. B}\ }\textbf {\bibinfo {volume}
  {64}},\ \bibinfo {pages} {195109} (\bibinfo {year} {2001})}\BibitemShut
  {NoStop}%
\bibitem [{\citenamefont {Metzner}\ \emph {et~al.}(2003)\citenamefont
  {Metzner}, \citenamefont {Rohe},\ and\ \citenamefont
  {Andergassen}}]{metzner-2003}%
  \BibitemOpen
  \bibfield  {author} {\bibinfo {author} {\bibfnamefont {W.}~\bibnamefont
  {Metzner}}, \bibinfo {author} {\bibfnamefont {D.}~\bibnamefont {Rohe}}, \
  and\ \bibinfo {author} {\bibfnamefont {S.}~\bibnamefont {Andergassen}},\
  }\href {\doibase 10.1103/PhysRevLett.91.066402} {\bibfield  {journal}
  {\bibinfo  {journal} {Phys. Rev. Lett.}\ }\textbf {\bibinfo {volume} {91}},\
  \bibinfo {pages} {066402} (\bibinfo {year} {2003})}\BibitemShut {NoStop}%
\bibitem [{\citenamefont {Dell'Anna}\ and\ \citenamefont
  {Metzner}(2006)}]{metzner_2006}%
  \BibitemOpen
  \bibfield  {author} {\bibinfo {author} {\bibfnamefont {L.}~\bibnamefont
  {Dell'Anna}}\ and\ \bibinfo {author} {\bibfnamefont {W.}~\bibnamefont
  {Metzner}},\ }\href {\doibase 10.1103/PhysRevB.73.045127} {\bibfield
  {journal} {\bibinfo  {journal} {Phys. Rev. B}\ }\textbf {\bibinfo {volume}
  {73}},\ \bibinfo {pages} {045127} (\bibinfo {year} {2006})}\BibitemShut
  {NoStop}%
\bibitem [{\citenamefont {Rech}\ \emph {et~al.}(2006)\citenamefont {Rech},
  \citenamefont {P\'epin},\ and\ \citenamefont {Chubukov}}]{rech-pepin-2006}%
  \BibitemOpen
  \bibfield  {author} {\bibinfo {author} {\bibfnamefont {J.}~\bibnamefont
  {Rech}}, \bibinfo {author} {\bibfnamefont {C.}~\bibnamefont {P\'epin}}, \
  and\ \bibinfo {author} {\bibfnamefont {A.~V.}\ \bibnamefont {Chubukov}},\
  }\href {\doibase 10.1103/PhysRevB.74.195126} {\bibfield  {journal} {\bibinfo
  {journal} {Phys. Rev. B}\ }\textbf {\bibinfo {volume} {74}},\ \bibinfo
  {pages} {195126} (\bibinfo {year} {2006})}\BibitemShut {NoStop}%
\bibitem [{\citenamefont {Maslov}\ and\ \citenamefont
  {Chubukov}(2010)}]{maslov}%
  \BibitemOpen
  \bibfield  {author} {\bibinfo {author} {\bibfnamefont {D.~L.}\ \bibnamefont
  {Maslov}}\ and\ \bibinfo {author} {\bibfnamefont {A.~V.}\ \bibnamefont
  {Chubukov}},\ }\href {\doibase 10.1103/PhysRevB.81.045110} {\bibfield
  {journal} {\bibinfo  {journal} {Phys. Rev. B}\ }\textbf {\bibinfo {volume}
  {81}},\ \bibinfo {pages} {045110} (\bibinfo {year} {2010})}\BibitemShut
  {NoStop}%
\bibitem [{\citenamefont {Lawler}\ and\ \citenamefont
  {Fradkin}(2007)}]{lawler-fradkin-2007}%
  \BibitemOpen
  \bibfield  {author} {\bibinfo {author} {\bibfnamefont {M.~J.}\ \bibnamefont
  {Lawler}}\ and\ \bibinfo {author} {\bibfnamefont {E.}~\bibnamefont
  {Fradkin}},\ }\href {\doibase 10.1103/PhysRevB.75.033304} {\bibfield
  {journal} {\bibinfo  {journal} {Phys. Rev. B}\ }\textbf {\bibinfo {volume}
  {75}},\ \bibinfo {pages} {033304} (\bibinfo {year} {2007})}\BibitemShut
  {NoStop}%
\bibitem [{\citenamefont {Chubukov}\ and\ \citenamefont
  {Khveshchenko}(2006)}]{khvesh2006}%
  \BibitemOpen
  \bibfield  {author} {\bibinfo {author} {\bibfnamefont {A.~V.}\ \bibnamefont
  {Chubukov}}\ and\ \bibinfo {author} {\bibfnamefont {D.~V.}\ \bibnamefont
  {Khveshchenko}},\ }\href {\doibase 10.1103/PhysRevLett.97.226403} {\bibfield
  {journal} {\bibinfo  {journal} {Phys. Rev. Lett.}\ }\textbf {\bibinfo
  {volume} {97}},\ \bibinfo {pages} {226403} (\bibinfo {year}
  {2006})}\BibitemShut {NoStop}%
\bibitem [{\citenamefont {Abanov}\ \emph {et~al.}(2001)\citenamefont {Abanov},
  \citenamefont {Chubukov},\ and\ \citenamefont {Finkel'stein}}]{acf}%
  \BibitemOpen
  \bibfield  {author} {\bibinfo {author} {\bibfnamefont {A.}~\bibnamefont
  {Abanov}}, \bibinfo {author} {\bibfnamefont {A.~V.}\ \bibnamefont
  {Chubukov}}, \ and\ \bibinfo {author} {\bibfnamefont {A.~M.}\ \bibnamefont
  {Finkel'stein}},\ }\href {http://stacks.iop.org/0295-5075/54/i=4/a=488}
  {\bibfield  {journal} {\bibinfo  {journal} {EPL (Europhysics Letters)}\
  }\textbf {\bibinfo {volume} {54}},\ \bibinfo {pages} {488} (\bibinfo {year}
  {2001})}\BibitemShut {NoStop}%
\bibitem [{\citenamefont {Abanov}\ \emph {et~al.}(2003)\citenamefont {Abanov},
  \citenamefont {Chubukov},\ and\ \citenamefont {Schmalian}}]{acs}%
  \BibitemOpen
  \bibfield  {author} {\bibinfo {author} {\bibfnamefont {A.}~\bibnamefont
  {Abanov}}, \bibinfo {author} {\bibfnamefont {A.~V.}\ \bibnamefont
  {Chubukov}}, \ and\ \bibinfo {author} {\bibfnamefont {J.}~\bibnamefont
  {Schmalian}},\ }\href {\doibase 10.1080/0001873021000057123} {\bibfield
  {journal} {\bibinfo  {journal} {Advances in Physics}\ }\textbf {\bibinfo
  {volume} {52}},\ \bibinfo {pages} {119} (\bibinfo {year} {2003})}\BibitemShut
  {NoStop}%
\bibitem [{\citenamefont {Vekhter}\ and\ \citenamefont
  {Chubukov}(2004)}]{vekhter2004}%
  \BibitemOpen
  \bibfield  {author} {\bibinfo {author} {\bibfnamefont {I.}~\bibnamefont
  {Vekhter}}\ and\ \bibinfo {author} {\bibfnamefont {A.~V.}\ \bibnamefont
  {Chubukov}},\ }\href {\doibase 10.1103/PhysRevLett.93.016405} {\bibfield
  {journal} {\bibinfo  {journal} {Phys. Rev. Lett.}\ }\textbf {\bibinfo
  {volume} {93}},\ \bibinfo {pages} {016405} (\bibinfo {year}
  {2004})}\BibitemShut {NoStop}%
\bibitem [{\citenamefont {Wang}\ \emph {et~al.}(2016)\citenamefont {Wang},
  \citenamefont {Abanov}, \citenamefont {Altshuler}, \citenamefont
  {Yuzbashyan},\ and\ \citenamefont {Chubukov}}]{wang-abanov-2016}%
  \BibitemOpen
  \bibfield  {author} {\bibinfo {author} {\bibfnamefont {Y.}~\bibnamefont
  {Wang}}, \bibinfo {author} {\bibfnamefont {A.}~\bibnamefont {Abanov}},
  \bibinfo {author} {\bibfnamefont {B.~L.}\ \bibnamefont {Altshuler}}, \bibinfo
  {author} {\bibfnamefont {E.~A.}\ \bibnamefont {Yuzbashyan}}, \ and\ \bibinfo
  {author} {\bibfnamefont {A.~V.}\ \bibnamefont {Chubukov}},\ }\href {\doibase
  10.1103/PhysRevLett.117.157001} {\bibfield  {journal} {\bibinfo  {journal}
  {Phys. Rev. Lett.}\ }\textbf {\bibinfo {volume} {117}},\ \bibinfo {pages}
  {157001} (\bibinfo {year} {2016})}\BibitemShut {NoStop}%
\bibitem [{\citenamefont {L\"ohneysen}\ \emph {et~al.}(2007)\citenamefont
  {L\"ohneysen}, \citenamefont {Rosch}, \citenamefont {Vojta},\ and\
  \citenamefont {W\"olfle}}]{wolfle-rmp-2007}%
  \BibitemOpen
  \bibfield  {author} {\bibinfo {author} {\bibfnamefont {H.~v.}\ \bibnamefont
  {L\"ohneysen}}, \bibinfo {author} {\bibfnamefont {A.}~\bibnamefont {Rosch}},
  \bibinfo {author} {\bibfnamefont {M.}~\bibnamefont {Vojta}}, \ and\ \bibinfo
  {author} {\bibfnamefont {P.}~\bibnamefont {W\"olfle}},\ }\href {\doibase
  10.1103/RevModPhys.79.1015} {\bibfield  {journal} {\bibinfo  {journal} {Rev.
  Mod. Phys.}\ }\textbf {\bibinfo {volume} {79}},\ \bibinfo {pages} {1015}
  (\bibinfo {year} {2007})}\BibitemShut {NoStop}%
\bibitem [{\citenamefont {Senthil}(2008)}]{senthil-2008}%
  \BibitemOpen
  \bibfield  {author} {\bibinfo {author} {\bibfnamefont {T.}~\bibnamefont
  {Senthil}},\ }\href {\doibase 10.1103/PhysRevB.78.035103} {\bibfield
  {journal} {\bibinfo  {journal} {Phys. Rev. B}\ }\textbf {\bibinfo {volume}
  {78}},\ \bibinfo {pages} {035103} (\bibinfo {year} {2008})}\BibitemShut
  {NoStop}%
\bibitem [{\citenamefont {Metlitski}\ and\ \citenamefont
  {Sachdev}(2010{\natexlab{a}})}]{metlitski-sachdev-2010-1}%
  \BibitemOpen
  \bibfield  {author} {\bibinfo {author} {\bibfnamefont {M.~A.}\ \bibnamefont
  {Metlitski}}\ and\ \bibinfo {author} {\bibfnamefont {S.}~\bibnamefont
  {Sachdev}},\ }\href {\doibase 10.1103/PhysRevB.82.075127} {\bibfield
  {journal} {\bibinfo  {journal} {Phys. Rev. B}\ }\textbf {\bibinfo {volume}
  {82}},\ \bibinfo {pages} {075127} (\bibinfo {year}
  {2010}{\natexlab{a}})}\BibitemShut {NoStop}%
\bibitem [{\citenamefont {Metlitski}\ and\ \citenamefont
  {Sachdev}(2010{\natexlab{b}})}]{metlitski-sachdev-2010-2}%
  \BibitemOpen
  \bibfield  {author} {\bibinfo {author} {\bibfnamefont {M.~A.}\ \bibnamefont
  {Metlitski}}\ and\ \bibinfo {author} {\bibfnamefont {S.}~\bibnamefont
  {Sachdev}},\ }\href {\doibase 10.1103/PhysRevB.82.075128} {\bibfield
  {journal} {\bibinfo  {journal} {Phys. Rev. B}\ }\textbf {\bibinfo {volume}
  {82}},\ \bibinfo {pages} {075128} (\bibinfo {year}
  {2010}{\natexlab{b}})}\BibitemShut {NoStop}%
\bibitem [{\citenamefont {Jiang}\ \emph {et~al.}(2013)\citenamefont {Jiang},
  \citenamefont {Block}, \citenamefont {Mishmash}, \citenamefont {Garrison},
  \citenamefont {Sheng}, \citenamefont {Motrunich},\ and\ \citenamefont
  {Fisher}}]{jiang-fisher-2013}%
  \BibitemOpen
  \bibfield  {author} {\bibinfo {author} {\bibfnamefont {H.-C.}\ \bibnamefont
  {Jiang}}, \bibinfo {author} {\bibfnamefont {M.~S.}\ \bibnamefont {Block}},
  \bibinfo {author} {\bibfnamefont {R.~V.}\ \bibnamefont {Mishmash}}, \bibinfo
  {author} {\bibfnamefont {J.~R.}\ \bibnamefont {Garrison}}, \bibinfo {author}
  {\bibfnamefont {D.~N.}\ \bibnamefont {Sheng}}, \bibinfo {author}
  {\bibfnamefont {O.~I.}\ \bibnamefont {Motrunich}}, \ and\ \bibinfo {author}
  {\bibfnamefont {M.~P.~A.}\ \bibnamefont {Fisher}},\ }\href {\doibase
  10.1038/nature11732} {\bibfield  {journal} {\bibinfo  {journal} {Nature}\
  }\textbf {\bibinfo {volume} {493}},\ \bibinfo {pages} {39} (\bibinfo {year}
  {2013})}\BibitemShut {NoStop}%
\bibitem [{\citenamefont {Raghu}\ \emph {et~al.}(2015)\citenamefont {Raghu},
  \citenamefont {Torroba},\ and\ \citenamefont {Wang}}]{torroba-2015}%
  \BibitemOpen
  \bibfield  {author} {\bibinfo {author} {\bibfnamefont {S.}~\bibnamefont
  {Raghu}}, \bibinfo {author} {\bibfnamefont {G.}~\bibnamefont {Torroba}}, \
  and\ \bibinfo {author} {\bibfnamefont {H.}~\bibnamefont {Wang}},\ }\href
  {\doibase 10.1103/PhysRevB.92.205104} {\bibfield  {journal} {\bibinfo
  {journal} {Phys. Rev. B}\ }\textbf {\bibinfo {volume} {92}},\ \bibinfo
  {pages} {205104} (\bibinfo {year} {2015})}\BibitemShut {NoStop}%
\bibitem [{\citenamefont {Lee}(2009)}]{sslee-2009}%
  \BibitemOpen
  \bibfield  {author} {\bibinfo {author} {\bibfnamefont {S.-S.}\ \bibnamefont
  {Lee}},\ }\href {\doibase 10.1103/PhysRevB.80.165102} {\bibfield  {journal}
  {\bibinfo  {journal} {Phys. Rev. B}\ }\textbf {\bibinfo {volume} {80}},\
  \bibinfo {pages} {165102} (\bibinfo {year} {2009})}\BibitemShut {NoStop}%
\bibitem [{\citenamefont {Lee}(2018)}]{lee-review-2018}%
  \BibitemOpen
  \bibfield  {author} {\bibinfo {author} {\bibfnamefont {S.-S.}\ \bibnamefont
  {Lee}},\ }\href {\doibase 10.1146/annurev-conmatphys-031016-025531}
  {\bibfield  {journal} {\bibinfo  {journal} {Annual Review of Condensed Matter
  Physics}\ }\textbf {\bibinfo {volume} {9}},\ \bibinfo {pages} {227} (\bibinfo
  {year} {2018})},\ \Eprint
  {http://arxiv.org/abs/https://doi.org/10.1146/annurev-conmatphys-031016-025531}
  {https://doi.org/10.1146/annurev-conmatphys-031016-025531} \BibitemShut
  {NoStop}%
\bibitem [{\citenamefont {{Lederer}}\ \emph {et~al.}(2015)\citenamefont
  {{Lederer}}, \citenamefont {{Schattner}}, \citenamefont {{Berg}},\ and\
  \citenamefont {{Kivelson}}}]{lederer-2015}%
  \BibitemOpen
  \bibfield  {author} {\bibinfo {author} {\bibfnamefont {S.}~\bibnamefont
  {{Lederer}}}, \bibinfo {author} {\bibfnamefont {Y.}~\bibnamefont
  {{Schattner}}}, \bibinfo {author} {\bibfnamefont {E.}~\bibnamefont {{Berg}}},
  \ and\ \bibinfo {author} {\bibfnamefont {S.~A.}\ \bibnamefont {{Kivelson}}},\
  }\href {\doibase 10.1103/PhysRevLett.114.097001} {\bibfield  {journal}
  {\bibinfo  {journal} {Physical Review Letters}\ }\textbf {\bibinfo {volume}
  {114}},\ \bibinfo {eid} {097001} (\bibinfo {year} {2015})},\ \Eprint
  {http://arxiv.org/abs/1406.1193} {arXiv:1406.1193 [cond-mat.supr-con]}
  \BibitemShut {NoStop}%
\bibitem [{\citenamefont {{Schattner}}\ \emph {et~al.}(2016)\citenamefont
  {{Schattner}}, \citenamefont {{Lederer}}, \citenamefont {{Kivelson}},\ and\
  \citenamefont {{Berg}}}]{schattner-2016}%
  \BibitemOpen
  \bibfield  {author} {\bibinfo {author} {\bibfnamefont {Y.}~\bibnamefont
  {{Schattner}}}, \bibinfo {author} {\bibfnamefont {S.}~\bibnamefont
  {{Lederer}}}, \bibinfo {author} {\bibfnamefont {S.~A.}\ \bibnamefont
  {{Kivelson}}}, \ and\ \bibinfo {author} {\bibfnamefont {E.}~\bibnamefont
  {{Berg}}},\ }\href {\doibase 10.1103/PhysRevX.6.031028} {\bibfield  {journal}
  {\bibinfo  {journal} {Physical Review X}\ }\textbf {\bibinfo {volume} {6}},\
  \bibinfo {eid} {031028} (\bibinfo {year} {2016})},\ \Eprint
  {http://arxiv.org/abs/1511.03282} {arXiv:1511.03282 [cond-mat.supr-con]}
  \BibitemShut {NoStop}%
\bibitem [{\citenamefont {Xu}\ \emph {et~al.}(2017)\citenamefont {Xu},
  \citenamefont {Sun}, \citenamefont {Schattner}, \citenamefont {Berg},\ and\
  \citenamefont {Meng}}]{xu-sun-2017}%
  \BibitemOpen
  \bibfield  {author} {\bibinfo {author} {\bibfnamefont {X.~Y.}\ \bibnamefont
  {Xu}}, \bibinfo {author} {\bibfnamefont {K.}~\bibnamefont {Sun}}, \bibinfo
  {author} {\bibfnamefont {Y.}~\bibnamefont {Schattner}}, \bibinfo {author}
  {\bibfnamefont {E.}~\bibnamefont {Berg}}, \ and\ \bibinfo {author}
  {\bibfnamefont {Z.~Y.}\ \bibnamefont {Meng}},\ }\href {\doibase
  10.1103/PhysRevX.7.031058} {\bibfield  {journal} {\bibinfo  {journal} {Phys.
  Rev. X}\ }\textbf {\bibinfo {volume} {7}},\ \bibinfo {pages} {031058}
  (\bibinfo {year} {2017})}\BibitemShut {NoStop}%
\bibitem [{\citenamefont {Klein}\ \emph {et~al.}(2020)\citenamefont {Klein},
  \citenamefont {Schattner}, \citenamefont {Berg},\ and\ \citenamefont
  {Chubukov}}]{klein-2020}%
  \BibitemOpen
  \bibfield  {author} {\bibinfo {author} {\bibfnamefont {A.}~\bibnamefont
  {Klein}}, \bibinfo {author} {\bibfnamefont {Y.}~\bibnamefont {Schattner}},
  \bibinfo {author} {\bibfnamefont {E.}~\bibnamefont {Berg}}, \ and\ \bibinfo
  {author} {\bibfnamefont {A.~V.}\ \bibnamefont {Chubukov}},\ }\href@noop {} {\
   (\bibinfo {year} {2020})},\ \Eprint {http://arxiv.org/abs/2003.09431}
  {arXiv:2003.09431 [cond-mat.str-el]} \BibitemShut {NoStop}%
\bibitem [{\citenamefont {Xu}\ \emph {et~al.}(2020)\citenamefont {Xu},
  \citenamefont {Klein}, \citenamefont {Sun}, \citenamefont {Chubukov},\ and\
  \citenamefont {Meng}}]{xu-klein-2020}%
  \BibitemOpen
  \bibfield  {author} {\bibinfo {author} {\bibfnamefont {X.~Y.}\ \bibnamefont
  {Xu}}, \bibinfo {author} {\bibfnamefont {A.}~\bibnamefont {Klein}}, \bibinfo
  {author} {\bibfnamefont {K.}~\bibnamefont {Sun}}, \bibinfo {author}
  {\bibfnamefont {A.~V.}\ \bibnamefont {Chubukov}}, \ and\ \bibinfo {author}
  {\bibfnamefont {Z.~Y.}\ \bibnamefont {Meng}},\ }\href@noop {} {\  (\bibinfo
  {year} {2020})},\ \Eprint {http://arxiv.org/abs/2003.11573} {arXiv:2003.11573
  [cond-mat.str-el]} \BibitemShut {NoStop}%
\bibitem [{\citenamefont {Lee}\ and\ \citenamefont
  {Nagaosa}(1992)}]{lee-nagaosa-1992}%
  \BibitemOpen
  \bibfield  {author} {\bibinfo {author} {\bibfnamefont {P.~A.}\ \bibnamefont
  {Lee}}\ and\ \bibinfo {author} {\bibfnamefont {N.}~\bibnamefont {Nagaosa}},\
  }\href {\doibase 10.1103/PhysRevB.46.5621} {\bibfield  {journal} {\bibinfo
  {journal} {Phys. Rev. B}\ }\textbf {\bibinfo {volume} {46}},\ \bibinfo
  {pages} {5621} (\bibinfo {year} {1992})}\BibitemShut {NoStop}%
\bibitem [{\citenamefont {Halperin}\ \emph {et~al.}(1993)\citenamefont
  {Halperin}, \citenamefont {Lee},\ and\ \citenamefont {Read}}]{hlr-1993}%
  \BibitemOpen
  \bibfield  {author} {\bibinfo {author} {\bibfnamefont {B.~I.}\ \bibnamefont
  {Halperin}}, \bibinfo {author} {\bibfnamefont {P.~A.}\ \bibnamefont {Lee}}, \
  and\ \bibinfo {author} {\bibfnamefont {N.}~\bibnamefont {Read}},\ }\href
  {\doibase 10.1103/PhysRevB.47.7312} {\bibfield  {journal} {\bibinfo
  {journal} {Phys. Rev. B}\ }\textbf {\bibinfo {volume} {47}},\ \bibinfo
  {pages} {7312} (\bibinfo {year} {1993})}\BibitemShut {NoStop}%
\bibitem [{\citenamefont {Altshuler}\ \emph {et~al.}(1994)\citenamefont
  {Altshuler}, \citenamefont {Ioffe},\ and\ \citenamefont {Millis}}]{aim}%
  \BibitemOpen
  \bibfield  {author} {\bibinfo {author} {\bibfnamefont {B.~L.}\ \bibnamefont
  {Altshuler}}, \bibinfo {author} {\bibfnamefont {L.~B.}\ \bibnamefont
  {Ioffe}}, \ and\ \bibinfo {author} {\bibfnamefont {A.~J.}\ \bibnamefont
  {Millis}},\ }\href {\doibase 10.1103/PhysRevB.50.14048} {\bibfield  {journal}
  {\bibinfo  {journal} {Phys. Rev. B}\ }\textbf {\bibinfo {volume} {50}},\
  \bibinfo {pages} {14048} (\bibinfo {year} {1994})}\BibitemShut {NoStop}%
\bibitem [{\citenamefont {Nayak}\ and\ \citenamefont
  {Wilczek}(1994)}]{Nayak1994}%
  \BibitemOpen
  \bibfield  {author} {\bibinfo {author} {\bibfnamefont {C.}~\bibnamefont
  {Nayak}}\ and\ \bibinfo {author} {\bibfnamefont {F.}~\bibnamefont
  {Wilczek}},\ }\href {https://link.aps.org/doi/10.1103/PhysRevLett.93.016405}
  {\bibfield  {journal} {\bibinfo  {journal} {Nuclear Physics B}\ }\textbf
  {\bibinfo {volume} {417}},\ \bibinfo {pages} {359} (\bibinfo {year}
  {1994})}\BibitemShut {NoStop}%
\bibitem [{\citenamefont {Polchinski}(1994)}]{polchinski-1994}%
  \BibitemOpen
  \bibfield  {author} {\bibinfo {author} {\bibfnamefont {J.}~\bibnamefont
  {Polchinski}},\ }\href {\doibase
  https://doi.org/10.1016/0550-3213(94)90449-9} {\bibfield  {journal} {\bibinfo
   {journal} {Nuclear Physics B}\ }\textbf {\bibinfo {volume} {422}},\ \bibinfo
  {pages} {617 } (\bibinfo {year} {1994})}\BibitemShut {NoStop}%
\bibitem [{\citenamefont {Kim}\ \emph {et~al.}(1995{\natexlab{a}})\citenamefont
  {Kim}, \citenamefont {Lee},\ and\ \citenamefont {Wen}}]{kim-lee-wen-1995}%
  \BibitemOpen
  \bibfield  {author} {\bibinfo {author} {\bibfnamefont {Y.~B.}\ \bibnamefont
  {Kim}}, \bibinfo {author} {\bibfnamefont {P.~A.}\ \bibnamefont {Lee}}, \ and\
  \bibinfo {author} {\bibfnamefont {X.-G.}\ \bibnamefont {Wen}},\ }\href
  {\doibase 10.1103/PhysRevB.52.17275} {\bibfield  {journal} {\bibinfo
  {journal} {Phys. Rev. B}\ }\textbf {\bibinfo {volume} {52}},\ \bibinfo
  {pages} {17275} (\bibinfo {year} {1995}{\natexlab{a}})}\BibitemShut {NoStop}%
\bibitem [{\citenamefont {Kim}\ \emph {et~al.}(1995{\natexlab{b}})\citenamefont
  {Kim}, \citenamefont {Lee}, \citenamefont {Wen},\ and\ \citenamefont
  {Stamp}}]{kim-lee-wen-stamp-1995}%
  \BibitemOpen
  \bibfield  {author} {\bibinfo {author} {\bibfnamefont {Y.~B.}\ \bibnamefont
  {Kim}}, \bibinfo {author} {\bibfnamefont {P.~A.}\ \bibnamefont {Lee}},
  \bibinfo {author} {\bibfnamefont {X.-G.}\ \bibnamefont {Wen}}, \ and\
  \bibinfo {author} {\bibfnamefont {P.~C.~E.}\ \bibnamefont {Stamp}},\ }\href
  {\doibase 10.1103/PhysRevB.51.10779} {\bibfield  {journal} {\bibinfo
  {journal} {Phys. Rev. B}\ }\textbf {\bibinfo {volume} {51}},\ \bibinfo
  {pages} {10779} (\bibinfo {year} {1995}{\natexlab{b}})}\BibitemShut {NoStop}%
\bibitem [{\citenamefont {Jain}\ and\ \citenamefont
  {Anderson}(2009)}]{jain-anderson-2009}%
  \BibitemOpen
  \bibfield  {author} {\bibinfo {author} {\bibfnamefont {J.~K.}\ \bibnamefont
  {Jain}}\ and\ \bibinfo {author} {\bibfnamefont {P.~W.}\ \bibnamefont
  {Anderson}},\ }\href {\doibase 10.1073/pnas.0902901106} {\bibfield  {journal}
  {\bibinfo  {journal} {Proceedings of the National Academy of Sciences}\
  }\textbf {\bibinfo {volume} {106}},\ \bibinfo {pages} {9131} (\bibinfo {year}
  {2009})},\ \Eprint
  {http://arxiv.org/abs/https://www.pnas.org/content/106/23/9131.full.pdf}
  {https://www.pnas.org/content/106/23/9131.full.pdf} \BibitemShut {NoStop}%
\bibitem [{\citenamefont {Mross}\ \emph {et~al.}(2010)\citenamefont {Mross},
  \citenamefont {McGreevy}, \citenamefont {Liu},\ and\ \citenamefont
  {Senthil}}]{mross-mcgreevy-2010}%
  \BibitemOpen
  \bibfield  {author} {\bibinfo {author} {\bibfnamefont {D.~F.}\ \bibnamefont
  {Mross}}, \bibinfo {author} {\bibfnamefont {J.}~\bibnamefont {McGreevy}},
  \bibinfo {author} {\bibfnamefont {H.}~\bibnamefont {Liu}}, \ and\ \bibinfo
  {author} {\bibfnamefont {T.}~\bibnamefont {Senthil}},\ }\href {\doibase
  10.1103/PhysRevB.82.045121} {\bibfield  {journal} {\bibinfo  {journal} {Phys.
  Rev. B}\ }\textbf {\bibinfo {volume} {82}},\ \bibinfo {pages} {045121}
  (\bibinfo {year} {2010})}\BibitemShut {NoStop}%
\bibitem [{\citenamefont {Metlitski}\ \emph {et~al.}(2015)\citenamefont
  {Metlitski}, \citenamefont {Mross}, \citenamefont {Sachdev},\ and\
  \citenamefont {Senthil}}]{metlitski-mross-2015}%
  \BibitemOpen
  \bibfield  {author} {\bibinfo {author} {\bibfnamefont {M.~A.}\ \bibnamefont
  {Metlitski}}, \bibinfo {author} {\bibfnamefont {D.~F.}\ \bibnamefont
  {Mross}}, \bibinfo {author} {\bibfnamefont {S.}~\bibnamefont {Sachdev}}, \
  and\ \bibinfo {author} {\bibfnamefont {T.}~\bibnamefont {Senthil}},\ }\href
  {\doibase 10.1103/PhysRevB.91.115111} {\bibfield  {journal} {\bibinfo
  {journal} {Phys. Rev. B}\ }\textbf {\bibinfo {volume} {91}},\ \bibinfo
  {pages} {115111} (\bibinfo {year} {2015})}\BibitemShut {NoStop}%
\bibitem [{\citenamefont {Dalidovich}\ and\ \citenamefont
  {Lee}(2013)}]{dalidovich-lee-2013}%
  \BibitemOpen
  \bibfield  {author} {\bibinfo {author} {\bibfnamefont {D.}~\bibnamefont
  {Dalidovich}}\ and\ \bibinfo {author} {\bibfnamefont {S.-S.}\ \bibnamefont
  {Lee}},\ }\href {\doibase 10.1103/PhysRevB.88.245106} {\bibfield  {journal}
  {\bibinfo  {journal} {Phys. Rev. B}\ }\textbf {\bibinfo {volume} {88}},\
  \bibinfo {pages} {245106} (\bibinfo {year} {2013})}\BibitemShut {NoStop}%
\bibitem [{\citenamefont {Damia}\ \emph {et~al.}(2019)\citenamefont {Damia},
  \citenamefont {Kachru}, \citenamefont {Raghu},\ and\ \citenamefont
  {Torroba}}]{raghu-torroba-2019}%
  \BibitemOpen
  \bibfield  {author} {\bibinfo {author} {\bibfnamefont {J.~A.}\ \bibnamefont
  {Damia}}, \bibinfo {author} {\bibfnamefont {S.}~\bibnamefont {Kachru}},
  \bibinfo {author} {\bibfnamefont {S.}~\bibnamefont {Raghu}}, \ and\ \bibinfo
  {author} {\bibfnamefont {G.}~\bibnamefont {Torroba}},\ }\href {\doibase
  10.1103/PhysRevLett.123.096402} {\bibfield  {journal} {\bibinfo  {journal}
  {Phys. Rev. Lett.}\ }\textbf {\bibinfo {volume} {123}},\ \bibinfo {pages}
  {096402} (\bibinfo {year} {2019})}\BibitemShut {NoStop}%
\bibitem [{\citenamefont {Gu}\ \emph {et~al.}(2017)\citenamefont {Gu},
  \citenamefont {Qi},\ and\ \citenamefont {Stanford}}]{gu-2017}%
  \BibitemOpen
  \bibfield  {author} {\bibinfo {author} {\bibfnamefont {Y.}~\bibnamefont
  {Gu}}, \bibinfo {author} {\bibfnamefont {X.-L.}\ \bibnamefont {Qi}}, \ and\
  \bibinfo {author} {\bibfnamefont {D.}~\bibnamefont {Stanford}},\ }\href
  {\doibase 10.1007/jhep05(2017)125} {\bibfield  {journal} {\bibinfo  {journal}
  {Journal of High Energy Physics}\ }\textbf {\bibinfo {volume} {2017}}
  (\bibinfo {year} {2017}),\ 10.1007/jhep05(2017)125}\BibitemShut {NoStop}%
\bibitem [{\citenamefont {Song}\ \emph {et~al.}(2017)\citenamefont {Song},
  \citenamefont {Jian},\ and\ \citenamefont {Balents}}]{balents-2017}%
  \BibitemOpen
  \bibfield  {author} {\bibinfo {author} {\bibfnamefont {X.-Y.}\ \bibnamefont
  {Song}}, \bibinfo {author} {\bibfnamefont {C.-M.}\ \bibnamefont {Jian}}, \
  and\ \bibinfo {author} {\bibfnamefont {L.}~\bibnamefont {Balents}},\ }\href
  {\doibase 10.1103/PhysRevLett.119.216601} {\bibfield  {journal} {\bibinfo
  {journal} {Phys. Rev. Lett.}\ }\textbf {\bibinfo {volume} {119}},\ \bibinfo
  {pages} {216601} (\bibinfo {year} {2017})}\BibitemShut {NoStop}%
\bibitem [{\citenamefont {Chowdhury}\ \emph {et~al.}(2018)\citenamefont
  {Chowdhury}, \citenamefont {Werman}, \citenamefont {Berg},\ and\
  \citenamefont {Senthil}}]{debanjan-prx}%
  \BibitemOpen
  \bibfield  {author} {\bibinfo {author} {\bibfnamefont {D.}~\bibnamefont
  {Chowdhury}}, \bibinfo {author} {\bibfnamefont {Y.}~\bibnamefont {Werman}},
  \bibinfo {author} {\bibfnamefont {E.}~\bibnamefont {Berg}}, \ and\ \bibinfo
  {author} {\bibfnamefont {T.}~\bibnamefont {Senthil}},\ }\href {\doibase
  10.1103/PhysRevX.8.031024} {\bibfield  {journal} {\bibinfo  {journal} {Phys.
  Rev. X}\ }\textbf {\bibinfo {volume} {8}},\ \bibinfo {pages} {031024}
  (\bibinfo {year} {2018})}\BibitemShut {NoStop}%
\bibitem [{\citenamefont {{Chowdhury}}\ and\ \citenamefont
  {{Berg}}(2019)}]{chowdhury-berg-2019}%
  \BibitemOpen
  \bibfield  {author} {\bibinfo {author} {\bibfnamefont {D.}~\bibnamefont
  {{Chowdhury}}}\ and\ \bibinfo {author} {\bibfnamefont {E.}~\bibnamefont
  {{Berg}}},\ }\href@noop {} {\bibfield  {journal} {\bibinfo  {journal} {arXiv
  e-prints}\ ,\ \bibinfo {eid} {arXiv:1908.02757}} (\bibinfo {year} {2019})},\
  \Eprint {http://arxiv.org/abs/1908.02757} {arXiv:1908.02757
  [cond-mat.str-el]} \BibitemShut {NoStop}%
\bibitem [{\citenamefont {{Wu}}\ \emph {et~al.}(2019)\citenamefont {{Wu}},
  \citenamefont {{Jian}},\ and\ \citenamefont {{Xu}}}]{xu-2019}%
  \BibitemOpen
  \bibfield  {author} {\bibinfo {author} {\bibfnamefont {X.-C.}\ \bibnamefont
  {{Wu}}}, \bibinfo {author} {\bibfnamefont {C.-M.}\ \bibnamefont {{Jian}}}, \
  and\ \bibinfo {author} {\bibfnamefont {C.}~\bibnamefont {{Xu}}},\ }\href@noop
  {} {\bibfield  {journal} {\bibinfo  {journal} {arXiv e-prints}\ ,\ \bibinfo
  {eid} {arXiv:1902.10154}} (\bibinfo {year} {2019})},\ \Eprint
  {http://arxiv.org/abs/1902.10154} {arXiv:1902.10154 [cond-mat.str-el]}
  \BibitemShut {NoStop}%
\bibitem [{\citenamefont {{Can}}\ and\ \citenamefont
  {{Franz}}(2019)}]{franz-2019}%
  \BibitemOpen
  \bibfield  {author} {\bibinfo {author} {\bibfnamefont {O.}~\bibnamefont
  {{Can}}}\ and\ \bibinfo {author} {\bibfnamefont {M.}~\bibnamefont
  {{Franz}}},\ }\href@noop {} {\bibfield  {journal} {\bibinfo  {journal} {arXiv
  e-prints}\ ,\ \bibinfo {eid} {arXiv:1903.00513}} (\bibinfo {year} {2019})},\
  \Eprint {http://arxiv.org/abs/1903.00513} {arXiv:1903.00513
  [cond-mat.str-el]} \BibitemShut {NoStop}%
\bibitem [{\citenamefont {Sachdev}\ and\ \citenamefont {Ye}(1993)}]{SY}%
  \BibitemOpen
  \bibfield  {author} {\bibinfo {author} {\bibfnamefont {S.}~\bibnamefont
  {Sachdev}}\ and\ \bibinfo {author} {\bibfnamefont {J.}~\bibnamefont {Ye}},\
  }\href {\doibase 10.1103/PhysRevLett.70.3339} {\bibfield  {journal} {\bibinfo
   {journal} {Phys. Rev. Lett.}\ }\textbf {\bibinfo {volume} {70}},\ \bibinfo
  {pages} {3339} (\bibinfo {year} {1993})}\BibitemShut {NoStop}%
\bibitem [{\citenamefont {Kitaev}(2015)}]{K}%
  \BibitemOpen
  \bibfield  {author} {\bibinfo {author} {\bibfnamefont {A.}~\bibnamefont
  {Kitaev}},\ }\href {http://online.kitp.ucsb.edu/online/entangled15/}
  {\bibfield  {journal} {\bibinfo  {journal} {Talks at KITP, University of
  California, Santa Barbara, Entanglement in Strongly-Correlated Quantum
  Matter}\ } (\bibinfo {year} {2015})}\BibitemShut {NoStop}%
\bibitem [{\citenamefont {Sachdev}(2015)}]{sachdev-2015}%
  \BibitemOpen
  \bibfield  {author} {\bibinfo {author} {\bibfnamefont {S.}~\bibnamefont
  {Sachdev}},\ }\href {\doibase 10.1103/PhysRevX.5.041025} {\bibfield
  {journal} {\bibinfo  {journal} {Phys. Rev. X}\ }\textbf {\bibinfo {volume}
  {5}},\ \bibinfo {pages} {041025} (\bibinfo {year} {2015})}\BibitemShut
  {NoStop}%
\bibitem [{\citenamefont {Maldacena}\ and\ \citenamefont
  {Stanford}(2016)}]{maldacena-2016}%
  \BibitemOpen
  \bibfield  {author} {\bibinfo {author} {\bibfnamefont {J.}~\bibnamefont
  {Maldacena}}\ and\ \bibinfo {author} {\bibfnamefont {D.}~\bibnamefont
  {Stanford}},\ }\href {\doibase 10.1103/physrevd.94.106002} {\bibfield
  {journal} {\bibinfo  {journal} {Physical Review D}\ }\textbf {\bibinfo
  {volume} {94}} (\bibinfo {year} {2016}),\
  10.1103/physrevd.94.106002}\BibitemShut {NoStop}%
\bibitem [{\citenamefont {Kitaev}\ and\ \citenamefont {Suh}(2018)}]{SYK3}%
  \BibitemOpen
  \bibfield  {author} {\bibinfo {author} {\bibfnamefont {A.}~\bibnamefont
  {Kitaev}}\ and\ \bibinfo {author} {\bibfnamefont {S.~J.}\ \bibnamefont
  {Suh}},\ }\href {\doibase 10.1007/JHEP05(2018)183} {\bibfield  {journal}
  {\bibinfo  {journal} {Journal of High Energy Physics}\ }\textbf {\bibinfo
  {volume} {2018}},\ \bibinfo {pages} {183} (\bibinfo {year}
  {2018})}\BibitemShut {NoStop}%
\bibitem [{\citenamefont {Altland}\ \emph {et~al.}(2019)\citenamefont
  {Altland}, \citenamefont {Bagrets},\ and\ \citenamefont
  {Kamenev}}]{atland2019}%
  \BibitemOpen
  \bibfield  {author} {\bibinfo {author} {\bibfnamefont {A.}~\bibnamefont
  {Altland}}, \bibinfo {author} {\bibfnamefont {D.}~\bibnamefont {Bagrets}}, \
  and\ \bibinfo {author} {\bibfnamefont {A.}~\bibnamefont {Kamenev}},\ }\href
  {\doibase 10.1103/PhysRevLett.123.106601} {\bibfield  {journal} {\bibinfo
  {journal} {Phys. Rev. Lett.}\ }\textbf {\bibinfo {volume} {123}},\ \bibinfo
  {pages} {106601} (\bibinfo {year} {2019})}\BibitemShut {NoStop}%
\bibitem [{\citenamefont {Gu}\ \emph {et~al.}(2020)\citenamefont {Gu},
  \citenamefont {Kitaev}, \citenamefont {Sachdev},\ and\ \citenamefont
  {Tarnopolsky}}]{gu-2020}%
  \BibitemOpen
  \bibfield  {author} {\bibinfo {author} {\bibfnamefont {Y.}~\bibnamefont
  {Gu}}, \bibinfo {author} {\bibfnamefont {A.}~\bibnamefont {Kitaev}}, \bibinfo
  {author} {\bibfnamefont {S.}~\bibnamefont {Sachdev}}, \ and\ \bibinfo
  {author} {\bibfnamefont {G.}~\bibnamefont {Tarnopolsky}},\ }\href {\doibase
  10.1007/JHEP02(2020)157} {\bibfield  {journal} {\bibinfo  {journal} {Journal
  of High Energy Physics}\ }\textbf {\bibinfo {volume} {2020}},\ \bibinfo
  {pages} {157} (\bibinfo {year} {2020})}\BibitemShut {NoStop}%
\bibitem [{\citenamefont {Davison}\ \emph {et~al.}(2017)\citenamefont
  {Davison}, \citenamefont {Fu}, \citenamefont {Georges}, \citenamefont {Gu},
  \citenamefont {Jensen},\ and\ \citenamefont {Sachdev}}]{davison-2017}%
  \BibitemOpen
  \bibfield  {author} {\bibinfo {author} {\bibfnamefont {R.~A.}\ \bibnamefont
  {Davison}}, \bibinfo {author} {\bibfnamefont {W.}~\bibnamefont {Fu}},
  \bibinfo {author} {\bibfnamefont {A.}~\bibnamefont {Georges}}, \bibinfo
  {author} {\bibfnamefont {Y.}~\bibnamefont {Gu}}, \bibinfo {author}
  {\bibfnamefont {K.}~\bibnamefont {Jensen}}, \ and\ \bibinfo {author}
  {\bibfnamefont {S.}~\bibnamefont {Sachdev}},\ }\href {\doibase
  10.1103/physrevb.95.155131} {\bibfield  {journal} {\bibinfo  {journal}
  {Physical Review B}\ }\textbf {\bibinfo {volume} {95}} (\bibinfo {year}
  {2017}),\ 10.1103/physrevb.95.155131}\BibitemShut {NoStop}%
\bibitem [{\citenamefont {Wang}(2020)}]{wang-2020}%
  \BibitemOpen
  \bibfield  {author} {\bibinfo {author} {\bibfnamefont {Y.}~\bibnamefont
  {Wang}},\ }\href {\doibase 10.1103/PhysRevLett.124.017002} {\bibfield
  {journal} {\bibinfo  {journal} {Phys. Rev. Lett.}\ }\textbf {\bibinfo
  {volume} {124}},\ \bibinfo {pages} {017002} (\bibinfo {year}
  {2020})}\BibitemShut {NoStop}%
\bibitem [{\citenamefont {Esterlis}\ and\ \citenamefont
  {Schmalian}(2019)}]{esterlis-2019}%
  \BibitemOpen
  \bibfield  {author} {\bibinfo {author} {\bibfnamefont {I.}~\bibnamefont
  {Esterlis}}\ and\ \bibinfo {author} {\bibfnamefont {J.}~\bibnamefont
  {Schmalian}},\ }\href {\doibase 10.1103/PhysRevB.100.115132} {\bibfield
  {journal} {\bibinfo  {journal} {Phys. Rev. B}\ }\textbf {\bibinfo {volume}
  {100}},\ \bibinfo {pages} {115132} (\bibinfo {year} {2019})}\BibitemShut
  {NoStop}%
\bibitem [{\citenamefont {{Hauck}}\ \emph {et~al.}(2019)\citenamefont
  {{Hauck}}, \citenamefont {{Klug}}, \citenamefont {{Esterlis}},\ and\
  \citenamefont {{Schmalian}}}]{esterlis-2019-2}%
  \BibitemOpen
  \bibfield  {author} {\bibinfo {author} {\bibfnamefont {D.}~\bibnamefont
  {{Hauck}}}, \bibinfo {author} {\bibfnamefont {M.~J.}\ \bibnamefont {{Klug}}},
  \bibinfo {author} {\bibfnamefont {I.}~\bibnamefont {{Esterlis}}}, \ and\
  \bibinfo {author} {\bibfnamefont {J.}~\bibnamefont {{Schmalian}}},\
  }\href@noop {} {\bibfield  {journal} {\bibinfo  {journal} {arXiv e-prints}\
  ,\ \bibinfo {eid} {arXiv:1911.04328}} (\bibinfo {year} {2019})},\ \Eprint
  {http://arxiv.org/abs/1911.04328} {arXiv:1911.04328 [cond-mat.str-el]}
  \BibitemShut {NoStop}%
\bibitem [{\citenamefont {Cao}\ \emph {et~al.}(2018{\natexlab{a}})\citenamefont
  {Cao}, \citenamefont {Fatemi}, \citenamefont {Demir}, \citenamefont {Fang},
  \citenamefont {Tomarken}, \citenamefont {Luo}, \citenamefont
  {Sanchez-Yamagishi}, \citenamefont {Watanabe}, \citenamefont {Taniguchi},
  \citenamefont {Kaxiras} \emph {et~al.}}]{cao-2018}%
  \BibitemOpen
  \bibfield  {author} {\bibinfo {author} {\bibfnamefont {Y.}~\bibnamefont
  {Cao}}, \bibinfo {author} {\bibfnamefont {V.}~\bibnamefont {Fatemi}},
  \bibinfo {author} {\bibfnamefont {A.}~\bibnamefont {Demir}}, \bibinfo
  {author} {\bibfnamefont {S.}~\bibnamefont {Fang}}, \bibinfo {author}
  {\bibfnamefont {S.~L.}\ \bibnamefont {Tomarken}}, \bibinfo {author}
  {\bibfnamefont {J.~Y.}\ \bibnamefont {Luo}}, \bibinfo {author} {\bibfnamefont
  {J.~D.}\ \bibnamefont {Sanchez-Yamagishi}}, \bibinfo {author} {\bibfnamefont
  {K.}~\bibnamefont {Watanabe}}, \bibinfo {author} {\bibfnamefont
  {T.}~\bibnamefont {Taniguchi}}, \bibinfo {author} {\bibfnamefont
  {E.}~\bibnamefont {Kaxiras}},  \emph {et~al.},\ }\href@noop {} {\bibfield
  {journal} {\bibinfo  {journal} {Nature}\ }\textbf {\bibinfo {volume} {556}},\
  \bibinfo {pages} {80} (\bibinfo {year} {2018}{\natexlab{a}})}\BibitemShut
  {NoStop}%
\bibitem [{\citenamefont {Cao}\ \emph {et~al.}(2018{\natexlab{b}})\citenamefont
  {Cao}, \citenamefont {Fatemi}, \citenamefont {Fang}, \citenamefont
  {Watanabe}, \citenamefont {Taniguchi}, \citenamefont {Kaxiras},\ and\
  \citenamefont {Jarillo-Herrero}}]{cao-2018-2}%
  \BibitemOpen
  \bibfield  {author} {\bibinfo {author} {\bibfnamefont {Y.}~\bibnamefont
  {Cao}}, \bibinfo {author} {\bibfnamefont {V.}~\bibnamefont {Fatemi}},
  \bibinfo {author} {\bibfnamefont {S.}~\bibnamefont {Fang}}, \bibinfo {author}
  {\bibfnamefont {K.}~\bibnamefont {Watanabe}}, \bibinfo {author}
  {\bibfnamefont {T.}~\bibnamefont {Taniguchi}}, \bibinfo {author}
  {\bibfnamefont {E.}~\bibnamefont {Kaxiras}}, \ and\ \bibinfo {author}
  {\bibfnamefont {P.}~\bibnamefont {Jarillo-Herrero}},\ }\href@noop {}
  {\bibfield  {journal} {\bibinfo  {journal} {Nature}\ }\textbf {\bibinfo
  {volume} {556}},\ \bibinfo {pages} {43} (\bibinfo {year}
  {2018}{\natexlab{b}})}\BibitemShut {NoStop}%
\bibitem [{\citenamefont {Coldea}\ and\ \citenamefont
  {Watson}(2018)}]{coldea-2018}%
  \BibitemOpen
  \bibfield  {author} {\bibinfo {author} {\bibfnamefont {A.~I.}\ \bibnamefont
  {Coldea}}\ and\ \bibinfo {author} {\bibfnamefont {M.~D.}\ \bibnamefont
  {Watson}},\ }\href@noop {} {\bibfield  {journal} {\bibinfo  {journal} {Annual
  Review of Condensed Matter Physics}\ }\textbf {\bibinfo {volume} {9}},\
  \bibinfo {pages} {125} (\bibinfo {year} {2018})}\BibitemShut {NoStop}%
\bibitem [{\citenamefont {Pan}\ \emph {et~al.}(2020)\citenamefont {Pan},
  \citenamefont {Wang},\ and\ \citenamefont {Meng}}]{pan-2020}%
  \BibitemOpen
  \bibfield  {author} {\bibinfo {author} {\bibfnamefont {G.}~\bibnamefont
  {Pan}}, \bibinfo {author} {\bibfnamefont {Y.}~\bibnamefont {Wang}}, \ and\
  \bibinfo {author} {\bibfnamefont {Z.~Y.}\ \bibnamefont {Meng}},\ }\href@noop
  {} {\  (\bibinfo {year} {2020})},\ \Eprint {http://arxiv.org/abs/2001.06586}
  {arXiv:2001.06586 [cond-mat.str-el]} \BibitemShut {NoStop}%
\bibitem [{\citenamefont {Bi}\ \emph {et~al.}(2017)\citenamefont {Bi},
  \citenamefont {Jian}, \citenamefont {You}, \citenamefont {Pawlak},\ and\
  \citenamefont {Xu}}]{cenke-syk-1}%
  \BibitemOpen
  \bibfield  {author} {\bibinfo {author} {\bibfnamefont {Z.}~\bibnamefont
  {Bi}}, \bibinfo {author} {\bibfnamefont {C.-M.}\ \bibnamefont {Jian}},
  \bibinfo {author} {\bibfnamefont {Y.-Z.}\ \bibnamefont {You}}, \bibinfo
  {author} {\bibfnamefont {K.~A.}\ \bibnamefont {Pawlak}}, \ and\ \bibinfo
  {author} {\bibfnamefont {C.}~\bibnamefont {Xu}},\ }\href {\doibase
  10.1103/PhysRevB.95.205105} {\bibfield  {journal} {\bibinfo  {journal} {Phys.
  Rev. B}\ }\textbf {\bibinfo {volume} {95}},\ \bibinfo {pages} {205105}
  (\bibinfo {year} {2017})}\BibitemShut {NoStop}%
\bibitem [{\citenamefont {Wu}\ \emph {et~al.}(2018)\citenamefont {Wu},
  \citenamefont {Chen}, \citenamefont {Jian}, \citenamefont {You},\ and\
  \citenamefont {Xu}}]{cenke-syk-2}%
  \BibitemOpen
  \bibfield  {author} {\bibinfo {author} {\bibfnamefont {X.}~\bibnamefont
  {Wu}}, \bibinfo {author} {\bibfnamefont {X.}~\bibnamefont {Chen}}, \bibinfo
  {author} {\bibfnamefont {C.-M.}\ \bibnamefont {Jian}}, \bibinfo {author}
  {\bibfnamefont {Y.-Z.}\ \bibnamefont {You}}, \ and\ \bibinfo {author}
  {\bibfnamefont {C.}~\bibnamefont {Xu}},\ }\href {\doibase
  10.1103/PhysRevB.98.165117} {\bibfield  {journal} {\bibinfo  {journal} {Phys.
  Rev. B}\ }\textbf {\bibinfo {volume} {98}},\ \bibinfo {pages} {165117}
  (\bibinfo {year} {2018})}\BibitemShut {NoStop}%
\bibitem [{\citenamefont {Patel}\ \emph {et~al.}(2018)\citenamefont {Patel},
  \citenamefont {Lawler},\ and\ \citenamefont {Kim}}]{patel-kim-2018}%
  \BibitemOpen
  \bibfield  {author} {\bibinfo {author} {\bibfnamefont {A.~A.}\ \bibnamefont
  {Patel}}, \bibinfo {author} {\bibfnamefont {M.~J.}\ \bibnamefont {Lawler}}, \
  and\ \bibinfo {author} {\bibfnamefont {E.-A.}\ \bibnamefont {Kim}},\ }\href
  {\doibase 10.1103/PhysRevLett.121.187001} {\bibfield  {journal} {\bibinfo
  {journal} {Phys. Rev. Lett.}\ }\textbf {\bibinfo {volume} {121}},\ \bibinfo
  {pages} {187001} (\bibinfo {year} {2018})}\BibitemShut {NoStop}%
\bibitem [{\citenamefont {Kim}\ \emph {et~al.}(2019)\citenamefont {Kim},
  \citenamefont {Klebanov}, \citenamefont {Tarnopolsky},\ and\ \citenamefont
  {Zhao}}]{zhao-prx-2019}%
  \BibitemOpen
  \bibfield  {author} {\bibinfo {author} {\bibfnamefont {J.}~\bibnamefont
  {Kim}}, \bibinfo {author} {\bibfnamefont {I.~R.}\ \bibnamefont {Klebanov}},
  \bibinfo {author} {\bibfnamefont {G.}~\bibnamefont {Tarnopolsky}}, \ and\
  \bibinfo {author} {\bibfnamefont {W.}~\bibnamefont {Zhao}},\ }\href {\doibase
  10.1103/PhysRevX.9.021043} {\bibfield  {journal} {\bibinfo  {journal} {Phys.
  Rev. X}\ }\textbf {\bibinfo {volume} {9}},\ \bibinfo {pages} {021043}
  (\bibinfo {year} {2019})}\BibitemShut {NoStop}%
\bibitem [{\citenamefont {{Cheipesh}}\ \emph {et~al.}(2019)\citenamefont
  {{Cheipesh}}, \citenamefont {{Pavlov}}, \citenamefont {{Scopelliti}},
  \citenamefont {{Tworzydlo}},\ and\ \citenamefont
  {{Gnezdilov}}}]{gnezdilov-2019}%
  \BibitemOpen
  \bibfield  {author} {\bibinfo {author} {\bibfnamefont {Y.}~\bibnamefont
  {{Cheipesh}}}, \bibinfo {author} {\bibfnamefont {A.~I.}\ \bibnamefont
  {{Pavlov}}}, \bibinfo {author} {\bibfnamefont {V.}~\bibnamefont
  {{Scopelliti}}}, \bibinfo {author} {\bibfnamefont {J.}~\bibnamefont
  {{Tworzydlo}}}, \ and\ \bibinfo {author} {\bibfnamefont {N.~V.}\ \bibnamefont
  {{Gnezdilov}}},\ }\href@noop {} {\bibfield  {journal} {\bibinfo  {journal}
  {arXiv e-prints}\ ,\ \bibinfo {eid} {arXiv:1910.00671}} (\bibinfo {year}
  {2019})},\ \Eprint {http://arxiv.org/abs/1910.00671} {arXiv:1910.00671
  [cond-mat.str-el]} \BibitemShut {NoStop}%
\bibitem [{\citenamefont {Wang}\ \emph {et~al.}(2020)\citenamefont {Wang},
  \citenamefont {Chudnovskiy}, \citenamefont {Gorsky},\ and\ \citenamefont
  {Kamenev}}]{wang-kamenev-2020}%
  \BibitemOpen
  \bibfield  {author} {\bibinfo {author} {\bibfnamefont {H.}~\bibnamefont
  {Wang}}, \bibinfo {author} {\bibfnamefont {A.~L.}\ \bibnamefont
  {Chudnovskiy}}, \bibinfo {author} {\bibfnamefont {A.}~\bibnamefont {Gorsky}},
  \ and\ \bibinfo {author} {\bibfnamefont {A.}~\bibnamefont {Kamenev}},\
  }\href@noop {} {\  (\bibinfo {year} {2020})},\ \Eprint
  {http://arxiv.org/abs/2002.11757} {arXiv:2002.11757 [cond-mat.str-el]}
  \BibitemShut {NoStop}%
\bibitem [{\citenamefont {{Kim}}\ \emph {et~al.}(2019)\citenamefont {{Kim}},
  \citenamefont {{Cao}},\ and\ \citenamefont {{Altman}}}]{kim-cao-altman-2019}%
  \BibitemOpen
  \bibfield  {author} {\bibinfo {author} {\bibfnamefont {J.}~\bibnamefont
  {{Kim}}}, \bibinfo {author} {\bibfnamefont {X.}~\bibnamefont {{Cao}}}, \ and\
  \bibinfo {author} {\bibfnamefont {E.}~\bibnamefont {{Altman}}},\ }\href@noop
  {} {\bibfield  {journal} {\bibinfo  {journal} {arXiv e-prints}\ ,\ \bibinfo
  {eid} {arXiv:1910.10173}} (\bibinfo {year} {2019})},\ \Eprint
  {http://arxiv.org/abs/1910.10173} {arXiv:1910.10173 [cond-mat.str-el]}
  \BibitemShut {NoStop}%
\bibitem [{\citenamefont {Kotliar}\ \emph {et~al.}(2006)\citenamefont
  {Kotliar}, \citenamefont {Savrasov}, \citenamefont {Haule}, \citenamefont
  {Oudovenko}, \citenamefont {Parcollet},\ and\ \citenamefont
  {Marianetti}}]{kotliar-2006}%
  \BibitemOpen
  \bibfield  {author} {\bibinfo {author} {\bibfnamefont {G.}~\bibnamefont
  {Kotliar}}, \bibinfo {author} {\bibfnamefont {S.~Y.}\ \bibnamefont
  {Savrasov}}, \bibinfo {author} {\bibfnamefont {K.}~\bibnamefont {Haule}},
  \bibinfo {author} {\bibfnamefont {V.~S.}\ \bibnamefont {Oudovenko}}, \bibinfo
  {author} {\bibfnamefont {O.}~\bibnamefont {Parcollet}}, \ and\ \bibinfo
  {author} {\bibfnamefont {C.~A.}\ \bibnamefont {Marianetti}},\ }\href
  {\doibase 10.1103/RevModPhys.78.865} {\bibfield  {journal} {\bibinfo
  {journal} {Rev. Mod. Phys.}\ }\textbf {\bibinfo {volume} {78}},\ \bibinfo
  {pages} {865} (\bibinfo {year} {2006})}\BibitemShut {NoStop}%
\bibitem [{\citenamefont {Georges}\ \emph {et~al.}(2001)\citenamefont
  {Georges}, \citenamefont {Parcollet},\ and\ \citenamefont
  {Sachdev}}]{gps-2001}%
  \BibitemOpen
  \bibfield  {author} {\bibinfo {author} {\bibfnamefont {A.}~\bibnamefont
  {Georges}}, \bibinfo {author} {\bibfnamefont {O.}~\bibnamefont {Parcollet}},
  \ and\ \bibinfo {author} {\bibfnamefont {S.}~\bibnamefont {Sachdev}},\ }\href
  {\doibase 10.1103/PhysRevB.63.134406} {\bibfield  {journal} {\bibinfo
  {journal} {Phys. Rev. B}\ }\textbf {\bibinfo {volume} {63}},\ \bibinfo
  {pages} {134406} (\bibinfo {year} {2001})}\BibitemShut {NoStop}%
\bibitem [{\citenamefont {Azeyanagi}\ \emph {et~al.}(2018)\citenamefont
  {Azeyanagi}, \citenamefont {Ferrari},\ and\ \citenamefont
  {Massolo}}]{azeyanagi-2018}%
  \BibitemOpen
  \bibfield  {author} {\bibinfo {author} {\bibfnamefont {T.}~\bibnamefont
  {Azeyanagi}}, \bibinfo {author} {\bibfnamefont {F.}~\bibnamefont {Ferrari}},
  \ and\ \bibinfo {author} {\bibfnamefont {F.~I.~S.}\ \bibnamefont {Massolo}},\
  }\href {\doibase 10.1103/PhysRevLett.120.061602} {\bibfield  {journal}
  {\bibinfo  {journal} {Phys. Rev. Lett.}\ }\textbf {\bibinfo {volume} {120}},\
  \bibinfo {pages} {061602} (\bibinfo {year} {2018})}\BibitemShut {NoStop}%
\bibitem [{\citenamefont {Patel}\ and\ \citenamefont
  {Sachdev}(2019)}]{patel-sachdev-2019}%
  \BibitemOpen
  \bibfield  {author} {\bibinfo {author} {\bibfnamefont {A.~A.}\ \bibnamefont
  {Patel}}\ and\ \bibinfo {author} {\bibfnamefont {S.}~\bibnamefont
  {Sachdev}},\ }\href {\doibase 10.1103/PhysRevLett.123.066601} {\bibfield
  {journal} {\bibinfo  {journal} {Phys. Rev. Lett.}\ }\textbf {\bibinfo
  {volume} {123}},\ \bibinfo {pages} {066601} (\bibinfo {year}
  {2019})}\BibitemShut {NoStop}%
\bibitem [{1fo()}]{1footnote}%
  \BibitemOpen
  \href@noop {} {}\bibinfo {note} {In a full self-consistent calculation with
  frequency-dependent $\Sigma (i\omega)$, $\Pi (i\Omega)$ is non-zero, but
  parameterically small compared to $m^2_0$.}\BibitemShut {Stop}%
\bibitem [{2fo()}]{2footnote}%
  \BibitemOpen
  \href@noop {} {}\bibinfo {note} {In this respect, NFL behavior in the Y-SYK
  model differs from that for dispersing itinerant fermions coupled to
  Landau-overdamped massless collective bosons, as there fermionic and bosonic
  self-energies do not change between frequencies, where $\Sigma (i\omega) >
  i\omega$, and $\Sigma (i\omega) < i\omega$.}\BibitemShut {Stop}%
\bibitem [{cod()}]{code}%
  \BibitemOpen
  \href@noop {} {}\bibinfo {howpublished} {The code for the iterative solution
  of the Schwinger-Dyson equations can be accessed at:
  \url{https://sites.google.com/site/yuxuanwang111/quantum-phase-transition-in-the-yukawa-syk-model?authuser=0}}\BibitemShut
  {NoStop}%
\bibitem [{wik()}]{wiki}%
  \BibitemOpen
  \href@noop {} {}\bibinfo {howpublished}
  {\url{https://en.wikipedia.org/wiki/Maxwell_construction}}\BibitemShut
  {NoStop}%
\bibitem [{\citenamefont {Baldwin}\ and\ \citenamefont
  {Swingle}(2019)}]{baldwin-swingle-2019}%
  \BibitemOpen
  \bibfield  {author} {\bibinfo {author} {\bibfnamefont {C.~L.}\ \bibnamefont
  {Baldwin}}\ and\ \bibinfo {author} {\bibfnamefont {B.}~\bibnamefont
  {Swingle}},\ }\href@noop {} {\  (\bibinfo {year} {2019})},\ \Eprint
  {http://arxiv.org/abs/1911.11865} {arXiv:1911.11865 [cond-mat.dis-nn]}
  \BibitemShut {NoStop}%
\bibitem [{\citenamefont {Tulipman}\ and\ \citenamefont
  {Berg}(2020)}]{tulipman-berg-2020}%
  \BibitemOpen
  \bibfield  {author} {\bibinfo {author} {\bibfnamefont {E.}~\bibnamefont
  {Tulipman}}\ and\ \bibinfo {author} {\bibfnamefont {E.}~\bibnamefont
  {Berg}},\ }\href@noop {} {\  (\bibinfo {year} {2020})},\ \Eprint
  {http://arxiv.org/abs/2004.03617} {arXiv:2004.03617 [cond-mat.str-el]}
  \BibitemShut {NoStop}%
\bibitem [{\citenamefont {Kruchkov}\ \emph {et~al.}(2020)\citenamefont
  {Kruchkov}, \citenamefont {Patel}, \citenamefont {Kim},\ and\ \citenamefont
  {Sachdev}}]{kruchkov-2020}%
  \BibitemOpen
  \bibfield  {author} {\bibinfo {author} {\bibfnamefont {A.}~\bibnamefont
  {Kruchkov}}, \bibinfo {author} {\bibfnamefont {A.~A.}\ \bibnamefont {Patel}},
  \bibinfo {author} {\bibfnamefont {P.}~\bibnamefont {Kim}}, \ and\ \bibinfo
  {author} {\bibfnamefont {S.}~\bibnamefont {Sachdev}},\ }\href {\doibase
  10.1103/PhysRevB.101.205148} {\bibfield  {journal} {\bibinfo  {journal}
  {Phys. Rev. B}\ }\textbf {\bibinfo {volume} {101}},\ \bibinfo {pages}
  {205148} (\bibinfo {year} {2020})}\BibitemShut {NoStop}%
\bibitem [{\citenamefont {Guo}\ \emph {et~al.}(2020)\citenamefont {Guo},
  \citenamefont {Gu},\ and\ \citenamefont {Sachdev}}]{guo-gu-sachdev-2020}%
  \BibitemOpen
  \bibfield  {author} {\bibinfo {author} {\bibfnamefont {H.}~\bibnamefont
  {Guo}}, \bibinfo {author} {\bibfnamefont {Y.}~\bibnamefont {Gu}}, \ and\
  \bibinfo {author} {\bibfnamefont {S.}~\bibnamefont {Sachdev}},\ }\href
  {\doibase https://doi.org/10.1016/j.aop.2020.168202} {\bibfield  {journal}
  {\bibinfo  {journal} {Annals of Physics}\ }\textbf {\bibinfo {volume}
  {418}},\ \bibinfo {pages} {168202} (\bibinfo {year} {2020})}\BibitemShut
  {NoStop}%
\bibitem [{\citenamefont {Luttinger}\ and\ \citenamefont
  {Ward}(1960)}]{luttinger-1960}%
  \BibitemOpen
  \bibfield  {author} {\bibinfo {author} {\bibfnamefont {J.~M.}\ \bibnamefont
  {Luttinger}}\ and\ \bibinfo {author} {\bibfnamefont {J.~C.}\ \bibnamefont
  {Ward}},\ }\href {\doibase 10.1103/PhysRev.118.1417} {\bibfield  {journal}
  {\bibinfo  {journal} {Phys. Rev.}\ }\textbf {\bibinfo {volume} {118}},\
  \bibinfo {pages} {1417} (\bibinfo {year} {1960})}\BibitemShut {NoStop}%
\bibitem [{\citenamefont {Altshuler}\ \emph {et~al.}(1998)\citenamefont
  {Altshuler}, \citenamefont {Chubukov}, \citenamefont {Dashevskii},
  \citenamefont {Finkel’stein},\ and\ \citenamefont {Morr}}]{altshuler-1998}%
  \BibitemOpen
  \bibfield  {author} {\bibinfo {author} {\bibfnamefont {B.~L.}\ \bibnamefont
  {Altshuler}}, \bibinfo {author} {\bibfnamefont {A.~V.}\ \bibnamefont
  {Chubukov}}, \bibinfo {author} {\bibfnamefont {A.}~\bibnamefont
  {Dashevskii}}, \bibinfo {author} {\bibfnamefont {A.~M.}\ \bibnamefont
  {Finkel’stein}}, \ and\ \bibinfo {author} {\bibfnamefont {D.~K.}\
  \bibnamefont {Morr}},\ }\href {\doibase 10.1209/epl/i1998-00164-y} {\bibfield
   {journal} {\bibinfo  {journal} {Europhysics Letters (EPL)}\ }\textbf
  {\bibinfo {volume} {41}},\ \bibinfo {pages} {401–406} (\bibinfo {year}
  {1998})}\BibitemShut {NoStop}%
\end{thebibliography}%

\end{document}